\documentclass[final,3p,times]{elsarticle}

\usepackage{amssymb}
\usepackage{amsmath}

\usepackage{tablefootnote}
\usepackage{booktabs}
\usepackage{multirow}
\usepackage{makecell}

\usepackage{siunitx}
\DeclareSIUnit{\torr}{Torr}
\DeclareSIUnit{\hour}{hr}
\DeclareSIUnit\angstrom{\text{Å}}
\newcommand{\s}[1]{_\mathrm{#1}}
\newcommand{\hd}{h\s{d}}
\newcommand{\hf}{h\s{f}}

\usepackage{url}

\usepackage{hyperref} 
\usepackage{cleveref} 

\journal{International Journal of Heat and Mass Transfer}

\begin{document}

\begin{frontmatter}

\title{Direct Modeling of Pore Size Evolution and Microcollapse in Lyophilization by Population Balance}

\author[che]{Isaac S. Wheeler} 
\author[che]{Vivek Narsimhan} 
\author[che,aae,bnc]{Alina A. Alexeenko} 
\author[pt]{Davide Fissore} 

\affiliation[che]{organization={Davidson School of Chemical Engineering, Purdue University},
            city={West Lafayette},
            state={IN},
            country={USA}}
\affiliation[aae]{organization={School of Aeronautics and Astronautics, Purdue University},
            city={West Lafayette},
            state={IN},
            country={USA}}
\affiliation[bnc]{organization={Birck Nanotechnology Center, Purdue University},
            city={West Lafayette},
            state={IN},
            country={USA}}
\affiliation[pt]{organization={Dipartimento di Scienza Applicata e Tecnologia, Politecnico di Torino},
            city={Torino},
            state={TO},
            country={Italy}}

\begin{abstract}
In the primary drying step of pharmaceutical freeze drying, it is well known that amorphous excipients are prone to ``microcollapse'' as they approach their glass transition temperature. Existing modeling approaches for this phenomenon generally entail either avoidance of this regime or an experimentally burdensome empirical assessment of the resulting drop in resistance to mass transfer. 
In this work, we approach the phenomenon head-on by developing a population balance model for the pore size distribution of a freeze-dried cake during microcollapse and relating the pore size to mass transfer resistance.
By treating the population balance with the method of moments and an interpolative closure, we keep the model's computational burden on par with the literature-standard model for primary drying.

For a literature example when mass transfer resistance clearly varies as a function of product temperature during drying, we fit the population balance model to cases with high and low product temperature and show that it can then simulate intermediate cases with equal accuracy.
The validity of the expression for mass transfer resistance is demonstrated with another literature data set, where pore size distributions were measured after the fact for a formulation without microcollapse; only a single parameter from this model needs to be fit to those cases to yield predictive accuracy in product temperature on par with post-hoc empirical assessment.
Finally, a sophisticated multi-experiment fit for one formulation across a range of conditions yields quantitative agreement in product temperatures, suggesting that the rate constant for microcollapse behavior may be a material parameter and independent of the precise porous geometry.
\end{abstract}

\begin{graphicalabstract}
\includegraphics[width=0.9\linewidth]{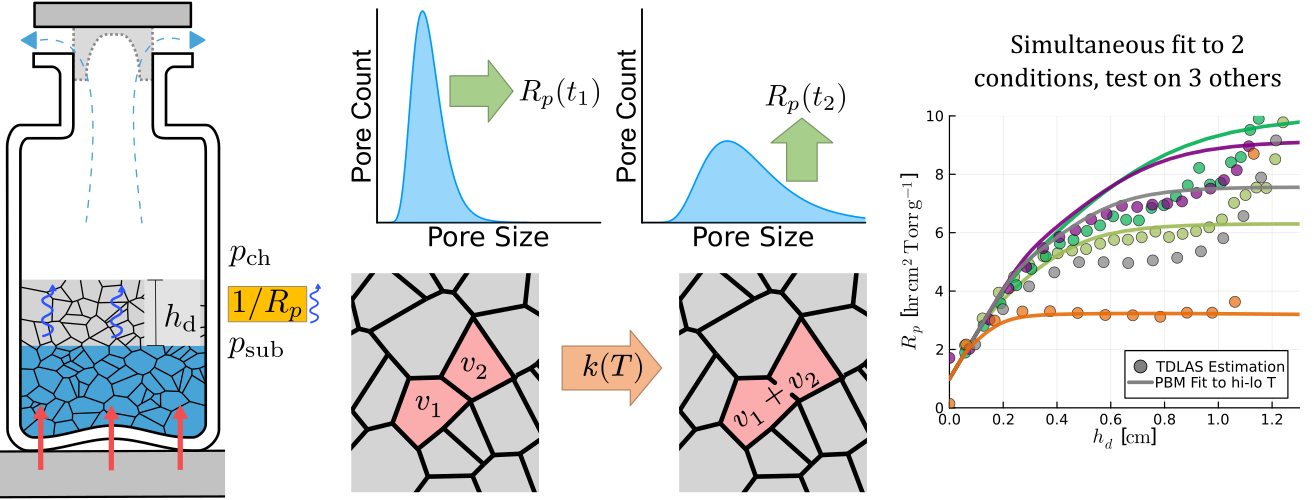}
\end{graphicalabstract}

\begin{highlights}
\item We give a population balance model for the open pore space during primary drying.  
\item This new model describes mass transfer variation across process conditions.
\item ``Microcollapse'' of pore structures corresponds to pore size changes. 
\item With measured pore size distributions, the new model needs less empirical fitting. 
\end{highlights}


\end{frontmatter}

\section{Introduction}

Freeze drying is an increasingly common step in drug product manufacturing, also called lyophilization because the resulting product can be rapidly dissolved (hence the Greek lyo, ``dissolve'', and phil, ``love'': wants to dissolve). In the pharmaceutical context, ``conventional'' freeze drying is carried out in a large vacuum chamber with temperature controlled shelves, on which vials containing drug product material are packed hexagonally. If loaded in a liquid state, the shelves are cooled from ambient temperature to \qty{-40}{\celsius} or lower to freeze the product; in other cases, already-frozen material is loaded onto precooled shelves. Once all material is loaded and frozen, the chamber is brought to a specified pressure, often in the vicinity of \qty{100}{\milli\torr} (\qty{13.3}{\pascal}), and shelves are brought to a specified temperature (typically between \qty{-40}{\celsius} and \qty{10}{\celsius}). Essentially pure ice sublimates away from the frozen material, leaving behind a porous phase with the drug product and excipients; this is called ``primary drying'', and its rate is governed by both the mass transfer resistance of the porous dried material and the thermal resistance from the shelf to the frozen product. ``Secondary drying'' follows, in which elevated shelf temperatures (up to \qty{40}{\celsius}) drive the porous material to a lower water content. 

An economic analysis of the cost of freeze drying \cite{strattaEconomicAnalysisFreezeDrying2020} reveals that the bulk of the cost is the amortized capital cost of the equipment, so by reducing the duration of primary drying (the longest step), we stand to cut the cost of this process. Even more importantly, however, as more lifesaving biologics are developed \cite{MerckModernaAnnounce}, the need for increased lyophilizer production capacity can be partially addressed by decreasing the length of the batch process. Biologics and other drugs are generally lyophilized because higher-temperature drying processes will have adverse effects on pharmaceutical safety or efficacy, and these temperature limits must also be respected by the freeze drying process as we make efforts to accelerate the process.  
The drying rate and product temperature are determined by shelf temperature and chamber pressure in a nonlinear way, so appropriate temperatures and pressures for drying should be chosen on the basis of a mathematical model.

Since its initial presentation in \cite{pikal_use_1985}, the vial-scale model for heat and mass transfer in the primary drying step of pharmaceutical lyophilization has been used with virtually no changes  \cite{tchessalov_practical_2023}, although now it is sometimes incorporated as a sub-model \cite{zadravec_towards_2025} or sampled across for probablistic approaches \cite{vanbillemont_optimization_2025, jameel_lyopronto_2023}. 
This model treats sublimation in a vial as pseudosteady, with 1D heating from the bottom to the sublimation front and 1D mass transfer through the dried product.
\citet{pikal_use_1985} originally stated that the porous product's mass transfer resistance $R_p$ is ``well represented by the empirical [form]''
\begin{equation}
    R_p = a_0 + \frac{a_1 h\s{d}}{1 + a_2 h\s{d}}  \label{eq:pikal-rp}
\end{equation}
which is a weakly nonlinear function of dried layer height $h\s{d}$.
This form has been used to empirically describe broad classes of behavior, particularly since the more predictable regimes tend to produce the desired uniform and pharmaceutically elegant products . 

$R_p$ can be measured in multiple ways. At the lab scale, it is simple to place thermocouples in a handful of vials, recording temperatures during lyophilization \citep{kuuRapidDeterminationDry2006}. Other PAT tools provide ways to measure batch-average mass flux, such as manometric temperature measurement \cite{milton1997evaluation}, tunable diode laser absorption spectroscopy (TDLAS) \cite{kuu_product_2011}, or taking differences between the inlet and outlet temperature of the shelf fluid \cite{authelin_simple_2024}. Measurements of either the product temperatures or mass flow rates can be inserted into Pikal' model to evaluate $h\s{d}(t)$ and $R_p(t)$, and typically \cref{eq:pikal-rp} is fit to this curve.  

The three empirical parameters $a_0$, $a_1$, and $a_2$ are generally found to vary with a multitude of process parameters. For example, this is clearly illustrated by \citet{lewis_characterizing_2010}, where for bare protein formulations overall curves for $R_p$ vary strongly with the overall solids content (5\% w/v or 20\% w/v), the drying conditions (conservative $T\s{sh}=\qty{-25}{\celsius} $ or aggressive $T\s{sh}=\qty{25}{\celsius}$), and even with proten identity (IgG, BSA, or lysozyme). Freezing behavior is also a key determinant of  mass transfer behavior, as attested by a large body of literature \cite{assegehegn_importance_2019}.

Some efforts have been made to tie \cref{eq:pikal-rp} to first principles, as derived by \citet{kuu_product_2011} and summarized by \citet{fissore_computer-aided_2015, wheeler_rarefied_2022}. That derivation approximates mass transfer in the porous dried layer as as Knudsen diffusion through capillaries with an effective radius $r_e$ and tortuosity $\tau$; to reproduce the denominator of $(1 + a_2)$ in \cref{eq:pikal-rp}, the effective radius is taken to vary linearly during drying as
\begin{equation}
    \frac{r_e}{\tau} = b_0 + b_1h\s{d} .
\end{equation}
That is, the nonlinear curve which is commonly observed in experiment (particularly for amorphous products \cite{pikal_use_1985, tchessalov_practical_2023}) implies that the effective radius of the capillaries is changing as drying proceeds: the imagined cylindrical capillaries grow \textit{radially}, along their entire length, as their length increases during drying. When \citet{konstantinidis_controlled_2011} analyzed $R_p$ this way, the effective radii grew by a factor of 3-5 across \qty{0.5}{\centi\meter} of dry layer height (see their Table 6). Capillaries increasing in diameter cannot be directly compared to images of dried porous structures. Thus, although this derivation provides a little more insight into the possible meaning of the parameters $a_1$ and $a_2$, the parameters $b_0$ and $b_1$ must still both be chosen to match experimental measurements and cannot be measured directly from porous structures. 

\citet{kosir_elucidating_2025} use two models which let $a_2=0$ and parameterize $a_1$ in terms of a permeability, in an effort to ground this fit parameter in first principles. Other authors, such as  \citet{witkiewicz_simulation_2010}, represent the resistance by an effective diffusivity through the porous layer, which has the same net effect of $R_p$ increasing linearly with dried layer height.

In this work, we develop a new approach for describing $R_p$ which makes it a more direct function of drying conditions; that is, as a function of the product temperature $T\s{f}$, which is coupled to $R_p$  but also depends on the shelf temperature $T\s{sh}$, the chamber pressure $p\s{ch}$, and the heat transfer coefficient $K_v$. We do so by accounting for how porous structures undergo microscale collapse during drying with a population balance. In doing so, we provide a relation for $R_p$ in terms of (the moments of) the pore size distribution. Prior studies \cite{kuu_product_2011, konstantinidis_controlled_2011, pisano_characterization_2017} have related $R_p$ to effective capillary radius, which is statistically equivalent to a first moment (or mean) of the pore size distribution. In other communities which study flow through porous media, it has been established that the \emph{second} moment of the pore size distribution also has fundamental connections to the permeability \cite{torquato_predicting_2020}. 

This article is organized as follows. In \cref{sec:pikal-statement}, we give the set of equations which we term the Pikal model, and in \cref{sec:model}, the new population balance is developed mathematically. We explain the experimental approach in \cref{sec:methods} and the algorithmic implementations in \cref{sec:numerics}. In \cref{sec:results}, the population balance model is fit to four sets of experiments: two from literature (\cref{sec:simtra-suc}, \cref{sec:kosir-dex})  and two newly conducted (\cref{sec:ful-dex}, \cref{sec:ful-suc}); these four sets may also grouped as two for amorphous formulations (\cref{sec:simtra-suc}, \cref{sec:ful-suc}) and two for crystalline formulations (\cref{sec:kosir-dex}, \cref{sec:ful-dex}). We discuss and present future directions in \cref{sec:discussion}, highlighting the model's practical utility (\cref{sec:utility}), the microcollapse rate constant (\cref{sec:results-temp-dep}, see also \cref{sec:model-temp-dep}), and rarefied flow regimes (\cref{sec:Kn}).
Finally, we conclude that the model we present actually describes porous microcollapse behavior, reducing the number of experiments necessary to design optimal freeze drying processes for amorphous formulations. 

\section{Literature-Standard Model} \label{sec:pikal-statement}
\begin{figure}
    \centering
    \includegraphics[width=0.5\linewidth]{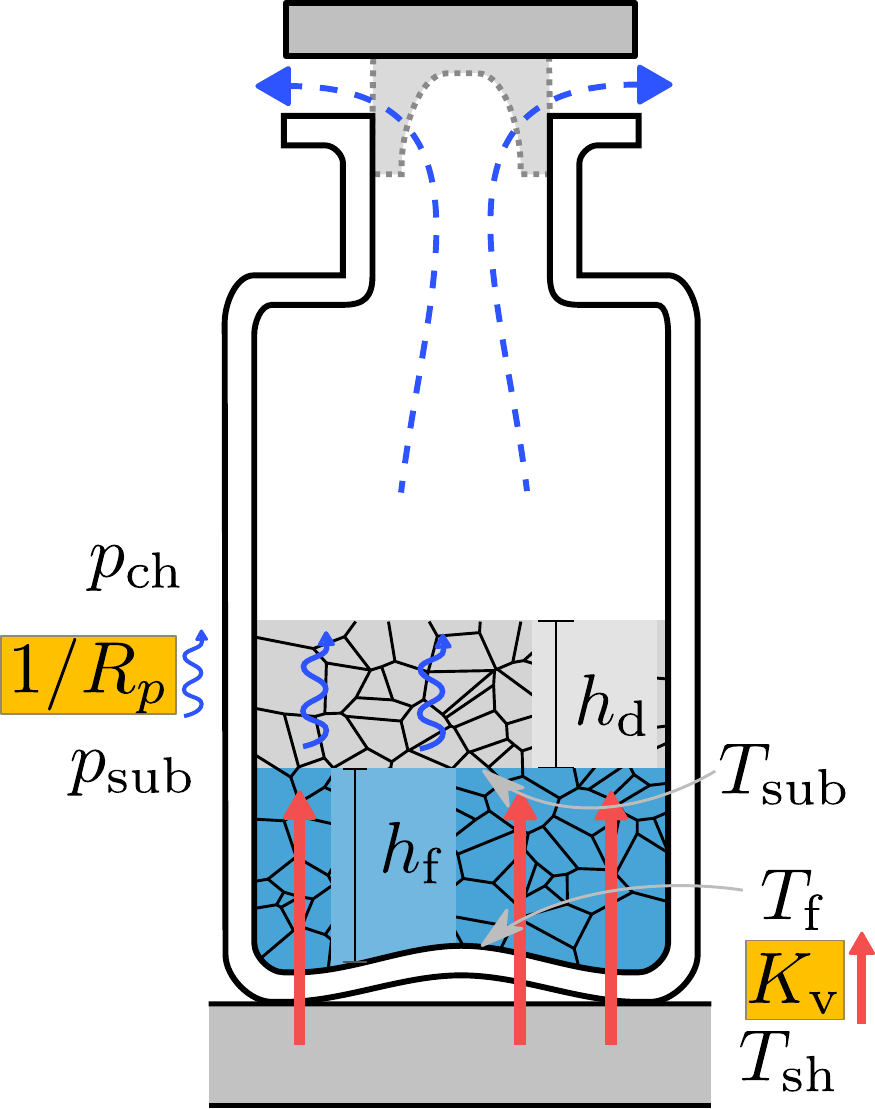}
    \caption{Schematic illustration of the Pikal model in a single vial.}
    \label{fig:vial-pikal}
\end{figure}

What we term the Pikal model \citep{pikal_use_1985} is stated as follows, and schematically illustrated in \cref{fig:vial-pikal}. The heat transfer $Q$ and mass transfer $\dot{m}$ are governed by heat transfer coefficient $K_v$ and product resistance $R_p$, respectively, and coupled by the latent heat of sublimation $\Delta H\s{sub}$:
\begin{align}
Q &= K_vA_v(T\s{sh}-T\s{f}) = \Delta H\s{sub} \dot{m} \label{eq:pikal-q}\\
\dot{m} &= \frac{A_p}{R_p}(p\s{sub}-p\s{ch}) 
\end{align}
with $A_v$ the ``vial area'' (computed from the vial's outer diameter) and $A_p$ the ``product area'' (computed from the vial's inner diameter). Shelf temperature $T\s{sh}$ and chamber pressure $p\s{ch}$ are directly-controlled process variables. The vapor pressure over ice $p\s{sub}$ is described by a simple correlation of similar form to the Clausius-Clapeyron equation, and is computed from the sublimation front temperature $T\s{sub}$. 
\begin{equation}
p\s{sub} = p\s{ref} \exp\left(- \frac{\Delta H\s{sub}}{RT\s{sub}}\right) \label{eq:psub} 
\end{equation}
$T\s{sub}$ is related to bottom product temperature $T\s{f}$ by 1D pseudosteady heat transfer through the ice, with thermal conductivity $\lambda\s{ice}$:
\begin{equation}
T\s{sub} = T\s{f} - \frac{Qh\s{f}}{A_p \lambda\s{ice}} . \label{eq:pikal-tsub}
\end{equation}
For simplicity, the frozen layer height $h\s{f}$ is initally taken equal to the liquid fill height (that is, neglecting expansion or shrinkage on freezing).
As drying proceeds, the frozen layer height decreases due to mass transfer, and the dry layer height $h\s{d}$ replaces it; this causes an increase in product resistance:
\begin{align}
    \frac{ dh\s{f}}{dt} &= \frac{\dot{m}}{\phi \rho\s{ice} A_p} \\
    h\s{d} &= h\s{f}(t=0) - h\s{f} = \frac{V\s{fill}}{A_p} - h\s{f}  \\
    R_p &= R_0 + \frac{A_1 h\s{d}}{1 + A_2 h\s{d}}
\end{align}
Ignoring the more rigorous expression originally given by \citet{pikal_use_1985}, the porosity $\phi$ is estimated as 
\begin{equation}
    \phi = \frac{\rho\s{water} - c\s{solid} }{\rho\s{water}}
\end{equation}
for a formulation concentration $c\s{solid}$ in mass units (e.g. \unit{\gram\per\milli\liter}).

Finally, at the vacuum conditions typical of lyophilization, the heat transfer coefficient $K_v$ depends on chamber pressure $p\s{ch}$ in a weakly nonlinear way that is justified by theoretical considerations \citep{pikal_use_1985, yoonUnderstandingHeatTransfer2022}: 
\begin{equation}
    K_v = K_C + \frac{K_P p\s{ch}}{1 + K_D p\s{ch}}. \label{eq:Kv}
\end{equation}
This expression holds in general for a given lyophilizer and vial type; if operating at a single pressure $p\s{ch}$, a single value of $K_v$ may be used directly without evaluating the parameters $K_C$, $K_P$, and $K_D$.

\section{Population Balance Modeling} \label{sec:model}

With the following model, we build on Pikal's model, particularly the 1D psuedosteady description of heat and mass transfer. We move beyond \cref{eq:pikal-rp} and instead connect $R_p$ to moments of the pore size distribution, which evolve over time according to a population balance model. 

\subsection{Pore Size Distribution}
We begin by assuming that the void space in the dried layer can be, somehow, partitioned into distinct pores with volumes denoted by $v$.
The details of how these pores are distinguished do not matter to the mathematical development; the important feature of this partitioning is that flow through a large pore is easier than through two smaller pores having a combined volume equal to the large pore. 
At a time $t$ during primary drying, the number of pores with a volume in the range $[v, v+dv]$ is denoted $n(v,t)dv$.
Summing together the volume of all of these pores (by integrating across all pore sizes) we find the total porous volume
\begin{equation}
    V(t) = \int_{v=0}^{v=\infty} vn(v,t)dv.
\end{equation}

\begin{figure}
    \centering
    \includegraphics[width=0.75\linewidth]{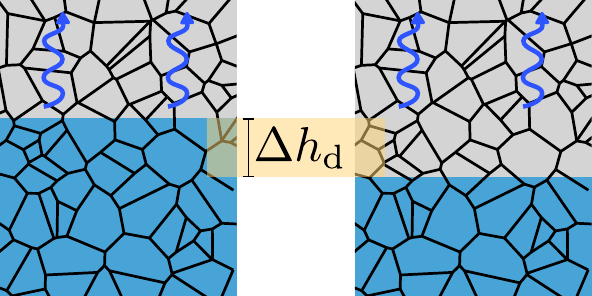}
    \caption{Schematic illustration: as primary drying proceeds, new pores are formed through which water vapor must flow.}
    \label{fig:add-pores}
\end{figure}

To simplify development, we assume that the porosity, or ratio of open pore volume to total volume, is constant throughout the dried cake, even when microcollapse occurs. 
We also therefore assume that the solid cake's intrinsic density does not change, and that the total volume of air and ice enclosed by the cake is constant (or in other words, that the cake does not shrink during primary drying).
The total pore volume is then equal to the dry layer's volume times porosity $\phi$:
\begin{equation}
    V(t) = \phi A\s{p} \hd.
\end{equation}
As mentioned in \cref{sec:pikal-statement}, we simply estimate the porosity $\phi$ by
$(\rho\s{water}-c\s{solids})/(\rho\s{water})$
so, for example, a 5\% w/v solution would have porosity of $(\qty{1}{\gram\per\milli\liter}-\qty{0.05}{\gram\per\milli\liter})/(\qty{1}{\gram\per\milli\liter}) = 0.95/1 = \num{0.95}$. 
As drying proceeds, the total porous volume increases---the sublimation of ice crystals leaves new open pores: 
\begin{equation}
    \frac{dV}{dt} = \phi A_p \frac{d\hd}{dt}. \label{eq:dVdt}
\end{equation}
This is schematically illustrated in \cref{fig:add-pores}, where a discrete change in height $\Delta h\s{d}$ is marked.

\subsection{Pore Formation and Collapse Phenomena}
The fundamental event that we will describe as microcollapse is the aggregation or coalescence of two pores into one larger pore. Using the language of chemical reactions, this may be thought of as an irreversible reaction:
\begin{equation}
    v_1 + v_2 \xrightarrow{k} (v_1+v_2)
\end{equation}
where $k$ is a rate constant describing how often or quickly this process occurs for a large collection of pores. In such a large collection, many discrete instances of these pore aggregation events may be thought of as taking place continuously. 

\begin{figure}
    \centering
    \includegraphics[width=0.75\linewidth]{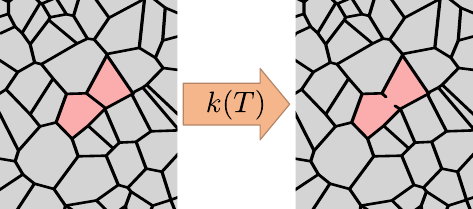}
    \caption{Schematic of a possible interpretation for single microcollapse events, which occurs with a rate constant $k(T)$.}
    \label{fig:microcollapse}
\end{figure}

Recall that the division of the pore space into discrete pores is not precisely defined; what matters for model utility is that, as far as flow through the structure is concerned, after a microcollapse event the two pores now behave as a single pore with correspondingly easier flow. This may mean that a new opening formed in a dividing wall, as schematically illustrated in \cref{fig:microcollapse}, or that an existing opening widened sufficiently to no longer obstruct the flow.

When one of these microcollapse events occurs, the numbers of pores with sizes $v_1$ and $v_2$ decrease, while the number of pores with size $(v_1+v_2)$ increases. 
Physically speaking, it is clear that to become a single pore, the two neighboring pores must be physically adjacent. 
Enforcing this condition mathematically could be done by defining a position-dependent pore distribution function, such as $n^*(z,v,t)$ such that
\begin{equation}
    n(v,t) = \int_{z=\hf}^{z=\hf+\hd} n^*(z,v,t)dz.
\end{equation}
However, in order to reduce the model complexity, we will track the pore distribution function irrespective of position and allow pores of all sizes and all positions to aggregate. This is mathematically equivalent to the ``well-mixed'' assumption in chemical reaction engineering, which is obviously impossible to realize for a set of stationary pores, but as we will show below this level of detail is still sufficient to lead to a useful model.

The rate of  ``births'' and ``deaths'' of pores of size $v$ can therefore be represented respectively by 
\begin{equation}
    \frac{1}{2}\int_{u=0}^{u=v} kn(u, t)n(v-u,t)du; \quad \int_{u=0}^{u=\infty}k n(v,t)n(u,t)du
\end{equation}
where the factor of 1/2 in the incrementing term prevents double-counting pore ``births'' (see section 3.3.2 of \citet{ramkrishna_population_2000}) and the rate constant $k$ has units of \unit{\per\second}. The factor of $n(v,t)$ can be factored out of the integral in the deaths term. If the rate constant $k$ is not a function of the two pore sizes, then it can be factored out of both integrals; this assumption was first utilized by Flory and demonstrated as valid in population balances describing condensing linear polymers \cite{floryKineticsPolyesterificationStudy1939a,floryMolecularSizeDistribution1941a,ziffKineticsPolymerization1980}; in other physical domains where these equations arise, this assumption might be called a mass-independent ``coagulation kernel''  \citep{hayakawaIrreversibleKineticCoagulations1987} or mass-independent merging rates for the Smoluchowski equations \citep{krapivskyGelationInputdrivenAggregation2024}.

The final term which must be accounted for to construct a master equation for the population of pores is the addition of new pores due to sublimation, as schematically shown in \cref{fig:add-pores}. Since new pores are added to the volume by the sublimation of ice crystals, the new pores should have a distribution of volumes similar to that of the ice crystals, which could in future be supplied by a first-principles model. This population of new pores will be described by a distribution function $n'$ such that $n'(z, v)dvdz$ is the number of pores having volume in the interval $[v, v+dv]$ which will be exposed by sublimation at the position $[z,z+dz]$. The rate of births of new pores at size $v$ and at a time $t$ where the current front position is $\hd$ is therefore
\begin{equation}
    n'(\hd,v)\frac{d\hd}{dt}.
\end{equation}
Assembling these birth and death terms into a master equation for our pore population, we have
\begin{equation}
    \frac{dn(v,t)}{dt} =  
    \frac{k}{2} \int_{u=0}^{u=v} n(u, t)n(v-u,t)du -
    kn(v,t)\int_{u=0}^{u=\infty} n(u,t)du +
    n'(\hd,v) \frac{d\hd}{dt}.
    \label{eq:master-equation}
\end{equation}

\subsection{Moment Generating Function and Transformed Master Equation}

The master equation (\cref{eq:master-equation}) could be discretized and solved in order to find the full distribution of pore sizes. However, in practice, the features of the distribution that we will most care about are features like the mean and variance, which are moments of the distribution. The method of moments can solve for these moments directly \cite{ramkrishna_population_2000}, reducing \cref{eq:master-equation} to a set of ordinary differential equations as follows.

The moment generating function $G(s,t)$ of our pore distribution function $n(v,t)$ is defined by
\begin{equation}
    G(s,t) = \int_{u=0}^{u=\infty} e^{su}n(u,t)du, \label{eq:moment-generating}
\end{equation}
and it has the useful property that  
\begin{equation}
   \left.\frac{\partial^k G}{\partial s^k}\right|_{s=0} = \int_0^\infty v^kn(v,t)dv=\mu_k(t)
\end{equation}
where $\mu_k$ denotes the $k$th moment of the pore distribution function. 
Since $n(v,t)$ is a pore \textit{count} rather than pore \textit{frequency}, these moments must be normalized by the zeroth moment to correspond to usual intuition.
For example, $\mu_0$ is the total number of pores, $\mu_1(t)=V(t)$ is the total pore volume discussed above, $\mu_1/\mu_0$ is the average pore size, and the variance of pore sizes is $\mu_2/\mu_0 - \mu_1^2/\mu_0^2$.

We can transform \cref{eq:master-equation} by multiplying by $e^{sv}$ and integrate over all $v$, as is done to the distribution function in \cref{eq:moment-generating}. With some manipulation to express all integrals in terms of the generating function $G$, we find the governing equation for the moment generating function:
\begin{equation}
    \frac{\partial}{\partial t}G(s,t) = \frac{k}{2} G^2(s,t) - kG(s,t)G(0,t) + G'(\hd, s)\frac{d\hd}{dt}, \label{eq:moment-governing-equation}
\end{equation}
 where $G'(\hd,s)=\int_0^\infty e^{sv}n'(\hd,v)dv$ is the moment generating function for the distribution of new pores.
 Differentiating this governing equation 0, 1, or 2 times with respect to $s$ then evaluating at $s=0$, we find the following differential equations for $\mu_0$, $\mu_1$, and $\mu_2$ respectively:
 \begin{align}
\frac{\partial}{\partial t} \mu_0 &=  G'(\hd, 0) \frac{d\hd}{dt} -\frac{k}{2} \mu_0^2,  \label{eq:mu0-pre}\\
\frac{\partial}{\partial t} \mu_1 &= \frac{\partial G'}{\partial s}(\hd,0)\frac{d\hd}{dt}, \\
\frac{\partial}{\partial t} \mu_2 &= \frac{\partial^2 G'}{\partial s^2}(\hd,0)\frac{d\hd}{dt} + k\mu_1^2. \label{eq:mu2-pre}
\end{align}
Further equations for higher moments could be derived in like manner.

The zeroth moment of the new pore distribution, $G'(\hd,0)$, is the number of new pores exposed per height of dried layer; with detailed information on the ice crystal size distribution, it could be given a dependence on $\hd$, but for the rest of the present work we will give it a constant value $NA_p$, where the factor $A_p$ normalizes area to remove vial size dependence. $N$ is thus the number of new pores per volume (or, equivalently, $\phi/N$ is the mean volume of new pores). Likewise, the second moment of the new pores could depend on $\hd$ but we will replace $\partial^2G'/\partial s^2(\hd,0)$ by $SA_p$. Recalling that the first moment of the pore size distribution is the total volume (\cref{eq:dVdt}), we can rewrite \crefrange{eq:mu0-pre}{eq:mu2-pre} as
\begin{align}
\frac{d\mu_0}{dt} &= NA_p \frac{d\hd}{dt} -\frac{k}{2} \mu_0^2 \label{eq:mu0-post}  \\
\frac{d\mu_1}{dt} &= \phi A_p \frac{d\hd}{dt} \label{eq:mu1-post}\\
\frac{d\mu_2}{dt} &= SA_p\frac{d\hd}{dt} + k\mu_1^2 \label{eq:mu2-post}
\end{align}
This very short set of equations thus has three empirical parameters to fit to experimental data: $N$, $S$, and $k$. 

We may motivate our use of the second moment as follows. 
Conceptually, the second moment adds more information about how many large pores are available for flow; larger pores not only admit more fluid but impose less resistance on fluid. If we consider a fixed number of straight cylindrical channels with a fixed total cross section area, so that adding area to the size of one pore must decrease the area of another pore, the flow \textit{resistance} will be \textit{maximized} if all the cylinders have the same area: two equal-sized channels will resist flow more than two channels where one is large and another is small. In this example, the first moment is fixed by fixing the number and total area, but the flow resistance can be adjusted by adjusting the variance (or second moment) of the pore sizes. 

\subsection{Relating Flow to Pore Distribution Moments }
To develop the simplest-possible relation of the flow (or, therefore, the mass transfer resistance $R_p$) to the moments of the pore size distribution, we examine the special case of a bundle of capillaries with free-molecular Knudsen diffusion (or with Hagen-Poiseuille viscous flow).

Let each capillary have the same length $l$ and a constant radius $r$. Let $f(r)dr$ be the number of capillaries with a radius in the interval $[r, r+dr]$. We would like to relate this to our volume-based pore counting $n(v,t)$, with the desired property that adding more length adds more pores without changing the distribution of radii. This suggests segmenting each capillary into volumes with radius $r$ and length $L<l$, so that a ``pore'' has volume $v=\pi r^2 L$ and the distributions are related by $f(r)dr = \frac{L}{l} n(v,t)dv$. The total pore count thus depends on segment length $L$, where the total volume does not. For example, a bundle of $N$ capillaries which share a single radius $r_0$ (i.e. $f(r) = N\delta(r-r_0) $ with the Dirac $\delta$) has a pore count $\mu_0 = Nl/L$, total volume $\mu_1 = N\pi r_0^2 l$, and second moment $\mu_2 = N\ \pi^2r_0^4lL$.

In the Knudsen flow regime, where water vapor is rarefied and particles interact more often with the capillary walls than with each other, the flux of vapor through a given capillary $J_K$ is proportional to the radius $r$. In a viscous flow regime, the Hagen-Poiseuille law states that the flux of water $J_V$ is proportional to $r^2$. Both $J_K$ and $J_V$ are also proportional to the pressure drop $\Delta p$ across the capillaries.

\begin{figure}
    \centering
    \includegraphics[width=0.8\linewidth]{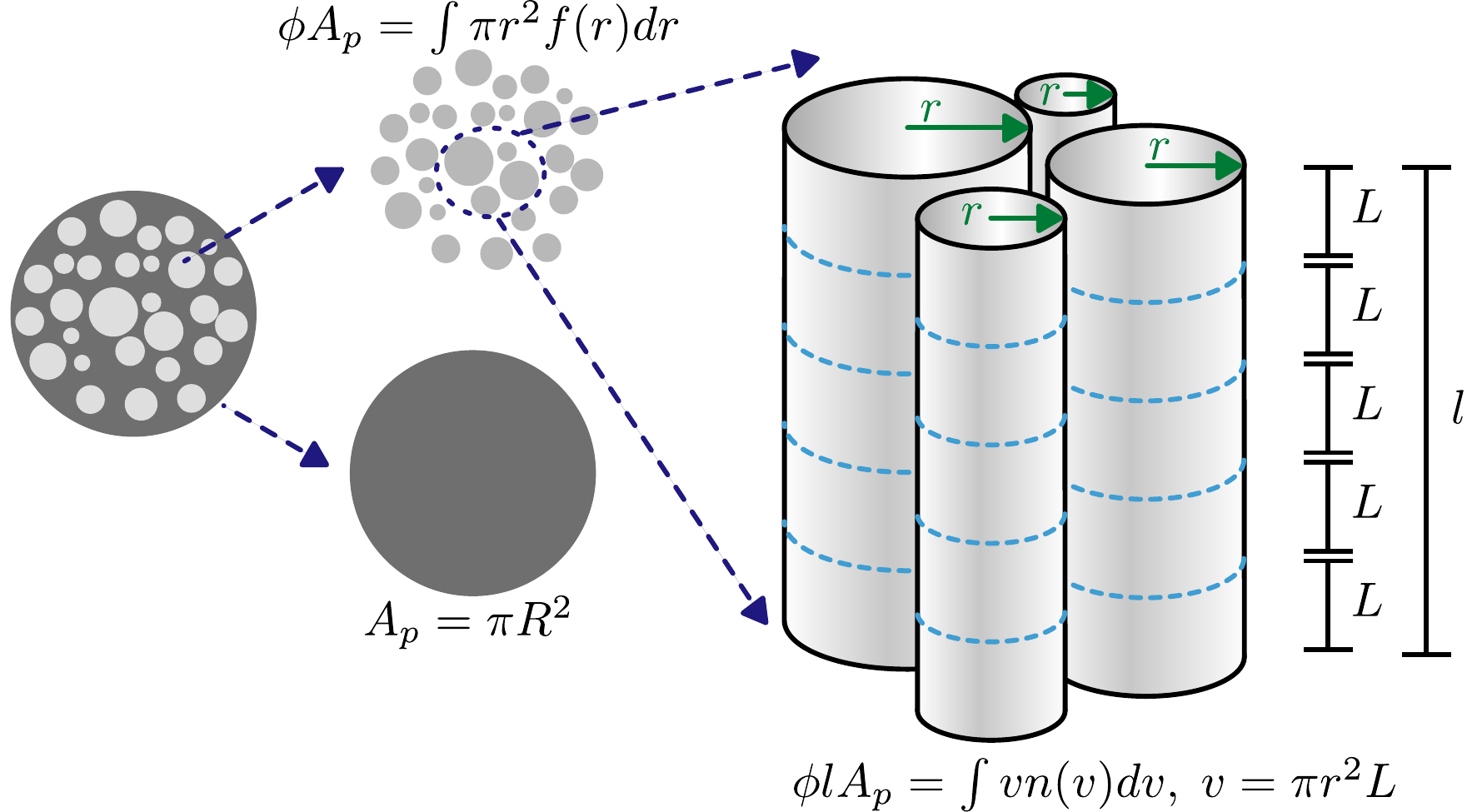}
    \caption{Schematic illustrating a bundle of capillaries with radius distribution $f(r)dr=\frac{L}{l}n(v)dv$, where $r$ is radius, $l$ is full length and $L<l$ is a length scale by which capillaries are segmented into smaller volumes.}
    \label{fig:capillaries}
\end{figure}
Since the total flow $\dot{m}$ is proportional to flux times area, multiply the flux for a given radius by capillary area and integrate across the distribution:
\begin{equation}
    \dot{m} = \int_0^\infty \pi r^2 J_if(r)dr =\int_0^\infty \frac{v}{L} J_i \frac{L}{l} n(v,t)dv=\int_0^\infty \frac{v}{l}J_in(v,t)dv. \label{eq:mass-from-flux}
\end{equation}
Plugging in expressions for $J_K$ and $J_V$ (given in \cref{eq:JK} and \cref{eq:JV} respectively), then solving for $R_p = A_p\Delta p/\dot{m}$,  we find the following two expressions for the mass transfer resistance, depending on the flow regime:
\begin{align}
R_K &= \frac{3\pi}{\sqrt{8}}\sqrt{L} \sqrt{\frac{R_uT}{M_w}} \frac{l\mu_1 }{\phi \mu_{3/2}} \label{eq:Rp-knudsen} \\
R_V &= 8\pi L \frac{\eta R_uT}{M_w p} \frac{l\mu_1 }{\phi \mu_2} 
\label{eq:Rp-viscous}
\end{align}
where $\eta$ is the dynamic viscosity of water vapor, $R_u$ is the universal gas constant, $M_w$ is the molar mass of water vapor,  $T$ and $p$ are the average temperature and pressure, and $\mu_{3/2} = \int_0^\infty v^{3/2}n(v,t)dv$  is a fractional moment of the pore space distribution. Since these expressions were derived for straight capillaries, we also introduce a dimensionless empirical prefactor $\tau$ for the tortuosity \citep{kuu_product_2011}. While the segment length $L$ is clearly connected to the way that pores are counted, there is at present no rational basis for identifying this length $L$ from arbitrary porous geometries, so we group it with the constant prefactors and the tortuosity into a single empirical constant $C$:
\begin{equation}
    C_K =\frac{3\pi}{\sqrt{8}} \tau \sqrt{L} , \quad C_V = 8\pi \tau L \label{eq:def-c}
\end{equation}
We will also add a nonzero but small mass transfer resistance, which prevents division by a zero mass transfer resistance at the start of drying. This resistance can be attributed to the stopper, plus any resistance due to a skin forming on the surface of the dried cake as is sometimes observed \cite{patel_lyophilized_2017}, so it will be denoted $R_s$.

In freeze drying, the Knudsen regime is much more likely than the viscous regime  \citep{kuu_product_2011, wheeler_rarefied_2022}. This is examined further in \cref{sec:Kn}. Therefore, in the rest of this work, we denote $C_K$ by $C$ and $R_K$ by $R_p$.

\subsection{Interpolative Closure for Fractional Moments}
Conveniently, the moment equations \cref{eq:mu0-post}-\cref{eq:mu2-post} do not themselves depend on any other moments than those being solved. However, to proceed, we need to evaluate $\mu_{3/2}=\int_0^\infty v^{3/2} n(v,t)dv$ as appears in \cref{eq:Rp-knudsen}. To do so, we turn to an interpolative closure (see \citet{frenklach_method_2002}'s section 7.1 for discussion of validity). This is done by Lagrange interpolation of the logarithms of the moments, with moment order as independent variable; with the three integer moments here, this yields
\begin{equation}
    \mu_{3/2} \approx \frac{\mu_1^{3/4} \mu_2^{3/8}}{\mu_0^{1/8}} . \label{eq:frac-mom} 
\end{equation}

\subsection{Temperature Dependence of Aggregation Rate Constant} \label{sec:model-temp-dep}

The pore aggregation parameter $k$ (with units \unit{\per\second}) is intended to represent the rate at which microcollapse events occur, which is when two distinct pores become a single pore for flow purposes. Empirical experience suggests that this is especially likely to occur for amorphous formulations when they approach their glass transition temperature. 
We expect $k$ to be a function of temperature, then, and tied to to the properties of the freeze-concentrated formulation.

It might not be unreasonable to additionally relate $k$ to the flow behavior in the cake, such as to the stresses on the cake due to water vapor flowing through it. Such a relation will likely depend in some sense on local pore sizes. 
However, this poses a mathematical difficulty: if the aggregation constant depends on pore size $v$, then the integrals performed in transforming the master equation \cref{eq:master-equation} into \cref{eq:moment-governing-equation} cannot be analytically evaluated as easily, and \cref{eq:mu0-post}--\cref{eq:mu2-post} will change as a consequence.
For simplicity, in the rest of this work, we will treat $k$ only as a function of temperature. 

In exploring the capabilities of this model, we consider two possible functional forms for $k$, as follows.
The first,
\begin{equation}
    k(T) = k\s{ref}\exp\left(\frac{K_1}{T}\exp\left[\frac{K_2}{T}\right]\right), \label{eq:MYEGA}
\end{equation}
is equivalent to the so-called MYEGA equation \cite{mauro_viscosity_2009}, which was developed to fit the relation between viscosity and temperature for glassy materials near their glass transition temperature. Insofar as microcollapse is a result of the freeze-concentrate being near its glass transition temperature, this would seem to be a natural way to describe the propensity for microcollapse, and further investigation could relate the empirical parameters here denoted $k\s{ref}$, $K_1$, and $K_2$ to glass transition temperatures and the configurational entropy or fragility of the glassy system \cite{mauro_viscosity_2009}.

The other form we will employ is
\begin{equation}
    k(T) = \frac{1}{2}[\tanh(x/2)+1](k\s{f}-k_0) + k_0 = \frac{k_0 + k\s{f} e^x}{1 + e^x} , \; x=\frac{T-T\s{act}}{\Theta}.
  \label{eq:sigmoid}
\end{equation}
We will call this the sigmoid form since it belongs to the class of sigmoid (``s-shaped'') functions; it has four parameters to determine empirically ($T\s{act}$, $\Theta$, $k_0$, and $k\s{f}$). For $T>T\s{act}$, this expression approaches an upper limit $k\s{f}$; for $T<T\s{act}$, it approaches $k_0$. The $T\s{act}$ thus should have some connection to the glass transition temperature. The value of $\Theta$ characterizes the sharpness of the transition from $k_0$ to $k\s{f}$, so for example with $\Theta=\qty{2}{\kelvin}$ the aggregation constant will go from nearly $k_0$ to nearly $k\s{f}$ across a temperature gap of about \qty{5}{\kelvin}.

Important to both of these equations is that, empirically, microcollapse accelerates rapidly in a fairly small range of temperatures---target temperatures \qtyrange{2}{5}{\celsius} below the critical temperature are typically considered safe for avoiding macroscale collapse \citep{tchessalov_practical_2023}, so the aggregation constant $k$ will likely go from negligible to significant across a change of only a few degrees Celsius.

\subsection{Summary of Population Balance Model}

The heat transfer behavior is taken to be exactly as in the traditional Pikal model, i.e. using \cref{eq:pikal-q} and \cref{eq:pikal-tsub} to relate the heat transfer $Q$ to shelf temperature $T\s{sh}$, the temperature at the bottom of the frozen layer $T\s{f}$ , and the temperature at the sublimation front $T\s{sub}$.
The pseudosteady 1D heat transfer again drives pseudosteady 1D mass transfer, with $R_p$ given instead by the moments of the pore size distribution (\cref{eq:Rp-knudsen}) :
\begin{align}
    \dot{m} = \frac{Q}{\Delta H\s{sub}} = \frac{A_p (p\s{sub} - p\s{ch}) }{R_s+R_p}. \\
    R_p = C \sqrt{\frac{R_u T\s{sub}}{M_W}} h\s{d} \frac{\mu_1^{1/4}\mu_0^{1/8}}{\mu_2^{3/8}},
\end{align}
where $R_s$ is a small, nonzero ``stopper resistance'', the empirical coefficient $C$ is given by \cref{eq:def-c}, and \cref{eq:frac-mom} for $\mu_{3/2}$ has been substituted and the results simplified.

As for the Pikal model, the height of the dried layer changes according to
\begin{equation}
    \frac{dh\s{d}}{dt} = \frac{\dot{m} }{A_p (\rho\s{water}-c\s{solids})}.
\end{equation}
The first moment $\mu_1$ (total pore volume) is given by $\phi A_p h\s{d}$. The total pore count $\mu_0$ and second moment of pore sizes $\mu_2$ are solved in time according to \cref{eq:mu0-post} and \cref{eq:mu2-post} respectively, with aggregation constant $k(T)$ given by either \cref{eq:MYEGA} or \cref{eq:sigmoid}, evaluated at $T\s{sub}$.

\subsection{Simplification under No-Microcollapse Conditions} \label{sec:no-microcollapse}
For crystalline formulations, it is common to observe that $R_p$ is strictly linear (i.e. $a_2=\qty{0}{\per\centi\meter}$ in \cref{eq:pikal-rp}). This is often taken to mean that no microcollapse occurs, which for this model implies that $k=\qty{0}{\per\second}$. In this case, \cref{eq:mu0-post} and \cref{eq:mu2-post} for $\mu_0$ and $\mu_2$ respectively simplify to $\mu_0(t)=NA_ph\s{d}(t)$ and $\mu_2(t)=SA_ph\s{d}(t)$. $R_p$ is then given by
\begin{equation}
    R_p  
    = C\sqrt{\frac{R_u T\s{sub}}{M_W}} h\s{d} \frac{\phi^{1/4}N^{1/8}}{S^{3/8}}.
\end{equation}
In this expression, since $C$, $N$, and $S$ are all directly multiplied, they cannot be separately estimated from data giving $R_p$ (or, equivalently, temperatures over time). However, if another means of estimating $N$ and $S$ is available, the number of fit parameters is reduced and instead only $C$ and stopper resistance $R_s$ remain as parameters which must be estimated from cycle data.

\section{Experimental Methods} \label{sec:methods}

\subsection{Literature Experimental Data}
We will compare our population balance model to two sets of experiments reported in the literature, for which in-process temperatures and pressures are reported. 

\citet{kosir_impact_2024} freeze dried 5-20\% w/v dextran solutions (which yield a crystalline product) with multiple freezing protocols and a single primary drying condition, and reported pore size distributions measured with scanning electron microscopy (SEM) imaging. The cases we analyze here, which they labeled A3, A5, A6, and B, are summarized in \cref{tab:kosir-cases}.
\citet{srinivasan_graphical_2023} freeze dried three formulations, of which we analyze Formulation 1, containing 1\% w/v BSA and 3\% w/v sucrose (which yields an amorphous product). They used a single freezing protocol and a range of primary drying conditions, of which we analyze the cases they labeled 1, 2, 4, 5, and 6, summarized in \cref{tab:srinivasan-cases}. In addition to temperatures and pressures, they reported mass flows measured by tunable diode laser absorption spectroscopy (TDLAS).

\begin{table}
    \centering
    \begin{tabular}{r|l l l l l l}
    \toprule
        \makecell{Case} & \makecell{Formulation, \\\% dextran} & \makecell{Vial \& fill} & \makecell{Freezing shelf \\ ramp, \unit{\celsius\per\minute}} & \makecell{Analyzed $T$ from Fig. S1}  & \makecell{$T\s{sh}$} & \makecell{$p\s{ch}$} \\
    \midrule
        A3 & 5\%  & \multirow{4}*{\makecell{\qty{3}{\milli\liter}, \\ 10R}} & \num{-0.5} & Darkest red 
         & \multirow{4}*{\qty{-5}{\celsius}} & \multirow{4}*{\qty{100}{\milli\torr}}\\
        A5 & 15\%  && \num{-0.5} & Lightest blue \textsuperscript{*} \\
        A6 & 20\%  && \num{-0.5} & Lightest green, until \qty{15}{\hour}   \\
        B  & 5\%  && \num{-0.2} & Lightest gray \textsuperscript{**}  \\

    \bottomrule
    \end{tabular}
    \caption{Summary of experiments from \citet{kosir_impact_2024} which we re-analyze. At beginning of primary drying, shelf temperature ramps from \qty{-45}{\celsius} at \qty{1}{\celsius\per\minute}. \\
    \textsuperscript{*}Other two reported trajectories stay cold for much longer than in other experiments, deviating from general trend. \\
    \textsuperscript{**}Other two reported trajectories have late nucleation times, which in their figure 4 correlates with very different pore size distributions.
    }
    \label{tab:kosir-cases}
\end{table}

\begin{table}
    \centering
    \begin{tabular}{r|l l l l}
    \toprule
        Case & Formulation & Vial \& fill & \makecell{$T\s{sh}$ (\unit{\celsius})} & \makecell{$p\s{ch}$ (\unit{\milli\torr})} \\
        \midrule
        1& \multirow{5}*{\makecell{1\% BSA,\\ 3\% sucrose}} &\multirow{5}*{ \makecell{\qty{5}{\milli\liter} fill, \\ \qty{10}{\milli\liter} vial}} & $-10$ & 100 \\
        2 &&& $-20$ & 50 \\
        4 &&& $-25$ & 100 \\
        5 &&& $-25$ & 60 \\
        6 &&& $-30$ & 100 \\
         \bottomrule
    \end{tabular}
    \caption{Summary of Formulation 1 cycles from \citet{srinivasan_graphical_2023}, their Table 8, which we re-analyze.}
    \label{tab:srinivasan-cases}
\end{table}

\subsection{Lyophilization of Sucrose- and Dextran-Based Solutions} \label{sec:ful-exp}
In addition to analyzing literature data, we performed experiments on sucrose and dextran solutions and examine the performance of the population balance model for these cases.

Solutions of 5\% w/v sucrose or 5\% w/v dextran were prepared by dissolving \qty{5}{\gram} of sucrose (purchased from Sigma Aldrich) or dextran (from \textit{Leuconostoc} spp., 40 kDa, purchased from Sigma Aldrich) respectively in \qty{100}{\milli\liter} of ultra-pure water from a Millipore water system (IQ 7000, Merck Millipore, Burlington, VT, USA). 10\% w/v solutions of sucrose were likewise prepared by dissolving \qty{10}{\gram} of sucrose.

We filled \qty{2}{\milli\liter} of solution into 10mL amber tubing vials, leading to a fill height of \qty{5.3}{\milli\meter}. A full complement of 19 vials were partially stoppered and placed in a MicroFD freeze dryer (Millrock Technology Inc., Kingston, NY) for lyophilization.

Unless otherwise noted, all samples were frozen by ramping the shelf temperature down from room temperature to \qty{-45}{\celsius} or \qty{-50}{\celsius} at \qty{-1}{\celsius\per\minute} and holding for 2-5 hours to allow full solidification. One batch of sucrose was frozen with controlled ice nucleation, specifically the ice fog method built in to the MicroFD dryer as FreezeBooster\textregistered; for this batch, shelf temperature was held at \qty{-7}{\celsius} for \qty{1}{\hour} to equilibrate all vials at this temperature, ice nucleation was triggered, then shelf temperature is immediately ramped down as for other batches. All primary drying was conducted at a chamber pressure of \qty{100}{\milli\torr}. For all cycles, when primary drying began, shelf temperature was ramped at \qty{0.5}{\celsius\per\minute} to a set point as given in \cref{tab:shelf-setpts}. Some time after Pirani pressure measurements approached capacitance manometer readings and reached a stable value, shelf set point was ramped to \qty{20}{\celsius} at \qty{0.5}{\celsius\per\minute} and held for 1-2 hours to carry out secondary drying before breaking vacuum and manual stoppering.

\begin{table}
    \centering
    \begin{tabular}{l|l}
        Formulation  &  Shelf set points, \unit{\celsius} \\ \midrule
        Sucrose, 5\% w/v & -20, -10, 0 \\
        Sucrose, 5\% w/v, CIN & -20 \\ 
        Sucrose, 10\% w/v & -30, -20 \\
        Dextran, 5\% w/v & -10, -5, 0, 5 \\
        \bottomrule
        
    \end{tabular}
    \caption{Summary of newly conducted experimental cycles, grouped by formulation, where each listed shelf temperature set point is a full experiment. For one cycle, we used controlled ice nucleation (CIN), which typically yields very different drying behavior by changing the freezing step.}
    \label{tab:shelf-setpts}
\end{table}

\subsection{SEM Imaging}

We removed lyophilized samples of sucrose and dextran from the vial by breaking the vial with a vise, after which samples were processed under a nitrogen atmosphere in a small nitrogen-fed box. Samples were sliced with razor blades along the vertical direction, as near to the center diameter as possible and again parallel to produce a slice with an approximate thickness of \qtyrange{2}{3}{\milli\meter}, exposing a vertical cross section through the center of the dried cake.
These slices were attached with double-sided tape to stubs, then sputter-coated with platinum for 30 seconds in a Quorum Q150T S sputter coater (2 M Instruments, Rome, Italy).

We used a Phenom XL benchtop scanning electron microscope to capture images at constant magnification, capturing a vertical sequence of overlapping images from the top to bottom near the center of each sample (with other images captured sporadically). The vertical sequence of images was stitched together by finding the relative offsets which minimized the L2 norm of grayscale value differences across the overlapping region, with a simple Julia script that can be found in the code repository \cite{isaacwheelerIckaserFulbrightPorousjlV0102026}.

\subsection{Determination of Heat Transfer Coefficient} \label{sec:fit-kv}
The heat transfer coefficient $K_v$ is well-known to depend on the specific lyophilizer and vial used, and is a function of pressure \cite{tchessalov_practical_2023}. For the experiments we carried out (as described in \cref{sec:ful-exp}), we measured this with the gravimetric method: filling \qty{2}{\milli\liter} of pure water rather than solid-bearing solution into the vials. Vials were numbered and weighed, then loaded into the dryer. After freezing at \qty{-45}{\celsius}, we ran an abbreviated 4-hour primary drying at $p\s{ch}=\qty{100}{\milli\torr}$ and $T\s{sh}=\qty{-20}{\celsius}$ , then broke vacuum, stoppered the vials manually, and weighed each vial immediately as well as after equilibrating at room temperature (to control for any condensation on the outside of cold vials). Thermocouple measurements in 2 vials during sublimation confirmed that, after shelf temperature reached its set point, ice temperatures remained within a degree of the frost point, i.e. $p\s{sub}(T\s{f})=p\s{ch}$, which is \qty{-39.6}{\celsius} for $p\s{ch}=\qty{100}{\milli\torr}$. With the average mass change due to sublimation  $ \Delta m = \frac{1}{n} \sum_i^n\Delta m_i$ across all vials $i$, $K_v$ can be computed by    
\begin{equation}
    K_v=\frac{\Delta H\s{sub} \Delta m}{A_v \int_0^{t\s{end}}[T\s{sh}(t)-T\s{f}(t)]dt}.
\end{equation}

\citet{srinivasan_graphical_2023} provide $K_v$ as a function of pressure for their lyophilizer and choice of vials, which yields $K_C=\qty{3.45}{\watt\per\meter\squared\per\kelvin}$, $K_P = \qty{193}{\watt\per\meter\squared\per\kelvin\per\torr}$, and $K_D=\qty{3.90}{\per\torr}$ for \cref{eq:Kv}.

Since \citet{kosir_impact_2024} do not report measured $K_v$, we derive the value by simultaneously fitting a single value of $K_v$ and distinct values of $R_p$ for each of their four cycles (that is, a 13-parameter fit for one $K_v$ and four each of $a_0$, $a_1$, and $a_2$) with the approach described below in \cref{sec:fits-to-data}

\section{Computational Approach} \label{sec:numerics}

\subsection{Identification of Primary Drying Time and Product Temperatures} \label{sec:exp-postprocess}

To fit models to experiment, we use both temperatures and times. As mentioned in \cite{wheelerMechanisticModelsExperimental2026} and discussed in \cref{sec:fits-to-data}, including drying time in the temperature fit improves the numerical performance of the fitting process.

\begin{figure}
    \centering
    \includegraphics[width=0.8\linewidth]{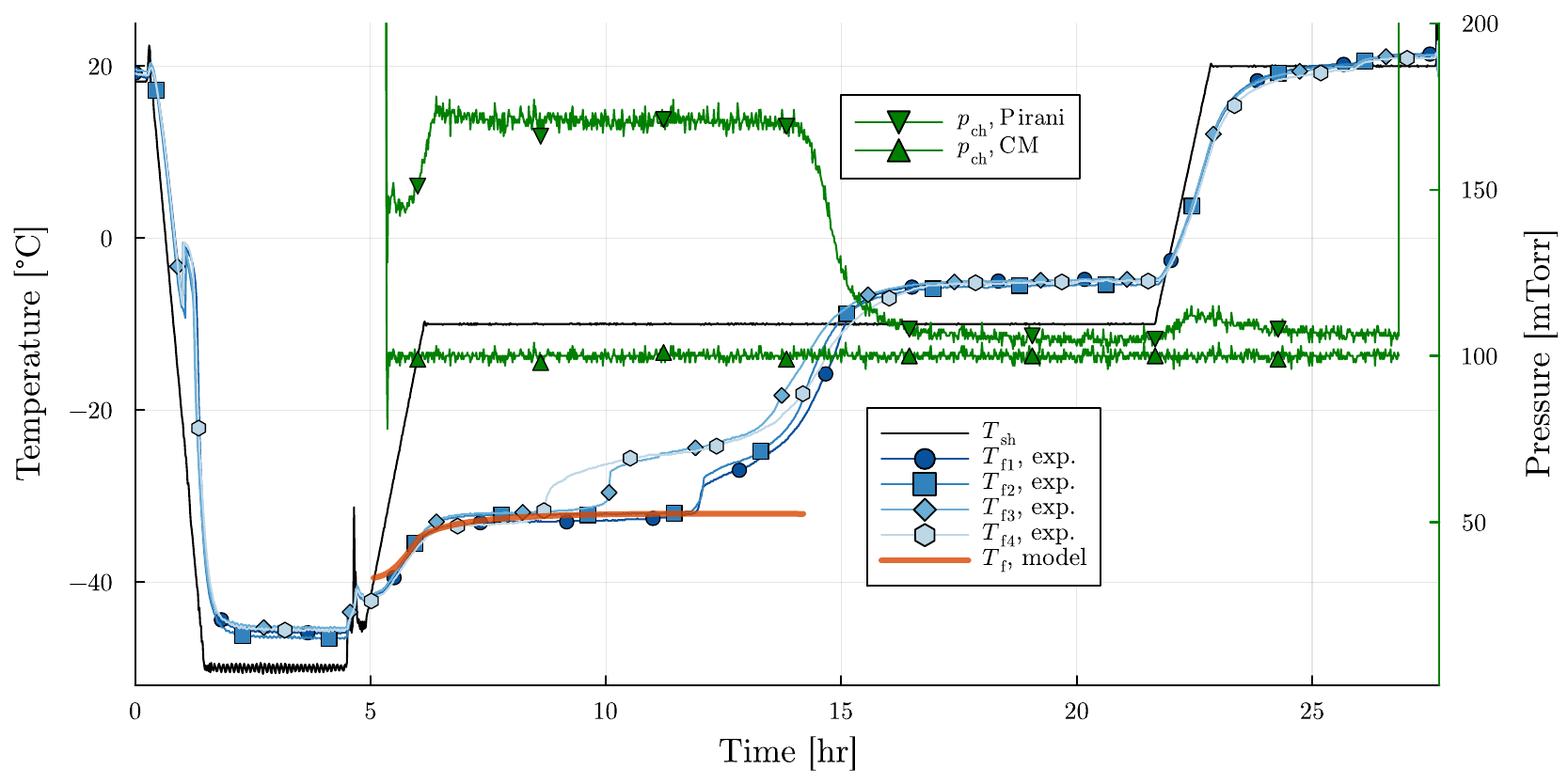}
    \caption{Typical lyophilization cycle data, as commonly represented with a single plot.
    The left axis here is temperature, used for the shelf and frozen product temperatures. The right axis is pressure, showing readings from a Pirani gauge and capacitance manometer in the chamber. $T\s{f1}$, $T\s{f2}$, etc. are thermocouple readings placed as nearly as possible in the bottom center of four frozen products, inside the vial. An example of temperature predictions with the Pikal model is shown in orange. The freezing step runs until about the \qty{5}{\hour} mark, and the lyophilizer moves from primary drying to secondary drying at about \qty{22}{\hour}.}
    \label{fig:example-cycle}
\end{figure}

\begin{figure}
    \centering
    \includegraphics[width=0.5\linewidth]{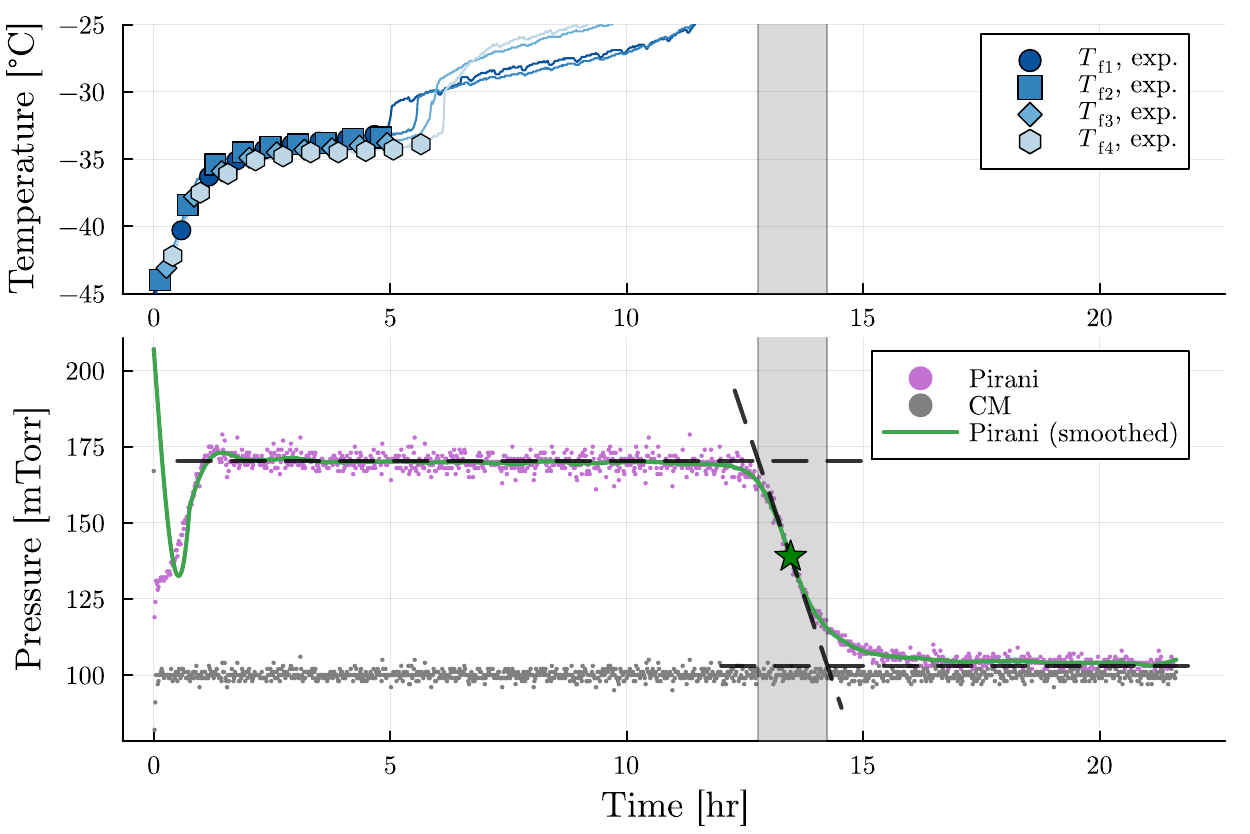}
    \caption{An example of the processing used to provide data for model fitting and validation. Temperature measurements are considered from the start of drying until the point where they curve upward from a simple trend (marked here with shaped markers), and process time is considered as a range from Pirani onset to offset.}
    \label{fig:example-time}
\end{figure}

An example of typical cycle data, with 4 thermocouples measuring product temperatures in different vials ($T\s{f1}$ through $T\s{f4}$), is shown in \cref{fig:example-cycle}, and the portions of that data that will be used for fitting primary drying models are shown in \cref{fig:example-time}. 

For the fitting process, we consider only the portions which correspond to primary drying. For convenience, time will be counted from the start of the primary drying phase. 
Temperatures for each frozen product are taken only up to the manually identified point where the temperature stops measure frozen layer temperature. This can be roughly identified where concavity changes from downward to upward and temperature rises rapidly, as seen in \cref{fig:example-cycle} between \qty{8}{\hour} and \qty{12}{\hour} when experimental measurements deviate from the model prediction. The time at which this rise occurs is dependent on exact thermocouple placement and exhibits significant variability, and may be an indication that the thermocouple is no longer within the frozen portion of the product \cite{wheelerMechanisticModelsExperimental2026}. In \cref{fig:example-time}, this is indicated by the end of markers in the temperature plot, near \qty{5}{\hour}. 

A number of PAT approaches have emerged in recent years for estimating the time at which primary drying is complete in a batch \cite{pisano_automatic_2022}. The ratio of pressures measured by the Pirani and capacitance manometer is easy to measure, and generally held to provide a reliable estimate of the end of drying; however, the dropoff of the curve as the ratio approaches 1 may span several hours and is dependent on the properties of the batch and equipment. This leads to some uncertainty in the exact point at which ice has actually finished sublimating.
For this reason, in this work we will show the endpoint of drying as a shaded region on temperature plots. This shaded region is constructed as follows. 

To identify the end of drying, first we identify the maximum downward slope of the Pirani curve, then construct a tangent line to the curve at that point (marked with a star in \cref{fig:example-time}). We identify this point and evaluate its slope with a Savitzky-Golay filter; a window width of 91 points and polynomial order 3 was found to give appropriate smoothing for this case.
The place where this tangent line intersects with the Pirani's maximum value during drying is sometimes referred to as the ``onset'' of Pirani-CM convergence, and the intersect with the Pirani gauge's minimum value after drying is likewise the ``offset'' of convergence. We take the range of time from onset to offset (\qtyrange{12.8}{14.2}{\hour} in \cref{fig:example-time}) as a bracketing interval for the end of primary drying: that is, the last ice in the product is likely to have sublimated sometime in that range.

\subsection{Fitting Mass Transfer Parameters to Experimental Data} \label{sec:fits-to-data}
Once $K_v$ is known, a few different approaches are available for fitting $R_p$ to experimental thermocouple measurements. One common approach, employed by LyoPRONTO \cite{shivkumar_lyopronto_2019} and others, is to evaluate the model equations with thermocouple measurements for $T\s{f}(t)$ and solve for $R_p$ and $h\s{d}$ as a function of time. Then, resulting curves for $R_p(h\s{d})$ are fit to \cref{eq:pikal-rp}.
In the Julia implementation \cite{isaacwheelerLyoHUBLyoProntojlV0452026}, this computation is available with the function \verb|calc_hRp_T|.

In our analysis of TDLAS data reported by \citet{srinivasan_graphical_2023}, we compute  $R_p$ in the same way it was computed there: for each measured time point $t_k$,
\begin{equation}
   R_p(t_k) =   A_p\frac{p\s{sub}(\bar T\s{f,k})-p\s{ch,k}}{\dot{m}_k}
   \label{eq:rp-tdlas}
\end{equation}
with $\bar T\s{f,k}$ the average product temperature across three thermocouples at time $k$, $p\s{ch,k}$ the chamber pressure at $t_k$ as measured by capacitance manometer, and $\dot m_k$ the TDLAS mass flow measurement. 
The dry layer height $\hd$ is computed as
\begin{equation}
    \hd(t_k)  =  h\s{f,0}\frac{ \sum_{i=1}^k \dot m_i}{\sum_{i=1}^n \dot m_i}
\end{equation}
where $n$ is the total number of measurements and negative values of $\dot m_i$ are clipped to zero.
This computation of $R_p$ tends to infinity at the end of drying because mass flow goes to zero while $p\s{sub}(\bar T\s{f,k})-p\s{ch}$ remains finite; this divergence is ignored since it can easily be identified manually, and any computed negative values (particularly at the beginning of primary drying) are ignored.

Computing $R_p(t_k)$ and $\hd(t_k)$ from experimentally measured $T\s{f}(t)$ often produces unphysical negative values at the beginning of the primary drying step, for example if $T\s{f} > T\s{sh}$ (so the shelf is cooling rather than heating the product) or $p\s{ch} > p\s{sub}(T\s{f})$ (drying is not proceeding), and these situations must be ignored since they produce negative or infinite values of $R_p$. In addition, experimental noise is transformed in a nonlinear way by this procedure, so statistical examination is more difficult.

To avoid these issues, we compare model output directly to experimental temperatures and find locally optimal fit parameters by solving a least squares problem, as follows.
Given some guess $\theta$ of the parameters to fit, the temperature residuals $E_{Ti}$ are constructed by solving the model equations to predict temperatures $\hat T\s{f}^i$ at time points $t^i$ where experimental measurements $T\s{f}^i$ are available:
\begin{equation}
    E_{Ti} = \hat T\s{f}^i - T\s{f}^i
\end{equation}
The total drying time is also considered as a residual in the least squares problem, which in practice helps to keep the model constrained to physically realizable regimes. For Pirani onset time $t\s{on}$ and offset time $t\s{off}$ (with midpoint $t\s{mid} = (t\s{on}+t\s{off})/2$, the residual $E_t$ for model end time $t\s{end}$ is
\begin{equation}
    E_t = \begin{cases}
        \qty{0}{\hour} & \text{if } t\s{on} \leq t\s{end} \leq t\s{off} \\
        t\s{end} - t\s{mid} & \text{otherwise}
    \end{cases}. \label{eq:time-resid}
\end{equation}
This residual is deliberately zero if the model predicted end time is between the Pirani onset and offset (\cref{sec:exp-postprocess}).

To considering both temperatures and time in the residual means a weighting factor must be chosen, which for this work is $w=\qty{0.1}{\kelvin\squared\per\hour\squared}$ (unless otherwise noted). This least squares problem is equivalent to minimizing a loss function $L$ over the set of fit parameters $\theta$:
\begin{equation}
    L(\theta) = wE_t^2 + \frac{1}{n}\sum_i^n E_{Ti}^2 \label{eq:loss}
\end{equation}
which has dimensions of \unit{\kelvin\squared}. This loss function is very similar in construction to Eq.~33 of \citet{wheelerMechanisticModelsExperimental2026}. To fit to multiple cycles at once, the loss function can be evaluated separately for each distinct experiment then summed. In practice, summing the loss functions this way enables new types of fits to experimental data, by letting the vector $\theta$ contain some parameters which are mapped to all experiments and some parameters which are mapped only to specific experiments. This novel functionality is facilitated by the open source package LyoPronto.jl \cite{isaacwheelerLyoHUBLyoProntojlV0452026}. 

In results below, goodness of fit is measured by reporting the root-mean-square error in temperature $\sqrt{\sum_i^n E_{Ti}^2/n} $ and the relative error in time $E_t /t\s{mid}$.  

\subsection{Numerical Implementation}

We formulate the Pikal model (stated below in \cref{sec:pikal-statement}) as a differential-algebraic equation (DAE) system with 1 ODE and 1 algebraic constraint; the new model introduced below is likewise formulated as a DAE with 3 ODEs and 1 algebraic constraint. Both are solved as an implicit ODE system with singular mass matrix in the Julia language with the \verb|Rodas4|  algorithm from \verb|OrdinaryDiffEqRosenbrock.jl| for DAE solution with \verb|ForwardDiff.jl| for automatic differentiation. The Pikal model solution and related utilities are available in the recently-developed package \verb|LyoPronto.jl| \cite{isaacwheelerLyoHUBLyoProntojlV0452026}, and we extend the same framework for the population balance model introduced here. 

To carry out parameter fitting, the least-squares problem defined in \cref{sec:fits-to-data} is solved. For multi-experiment fitting, it is treated as an \verb|OptimizationProblem| and solved with the \verb|Optim.jl| package, as wrapped in \\ \verb|OptimizationOptimJL.jl|, using the \verb|BFGS| algorithm with \verb|BackTracking| linesearch. For single-experiment fits, it was sometimes more convenient instead to formulate as a \verb|NonlinearLeastSquaresProblem| and solve with the \verb|GaussNewton| algorithm from \verb|NonlinearSolve.jl|. Carrying out this parameter fitting procedure in Julia allows us to make easy use of automatic differentiation (via \verb|ForwardDiff.jl|) to evaluate gradients of the residuals with respect to fit parameters without the need for finite differencing.

\section{Results} \label{sec:results}
\subsection{Literature Data: Amorphous Formulation with TDLAS Mass Flow Measurements} \label{sec:simtra-suc}

\begin{figure}
    \centering
    \includegraphics[width=0.95\linewidth]{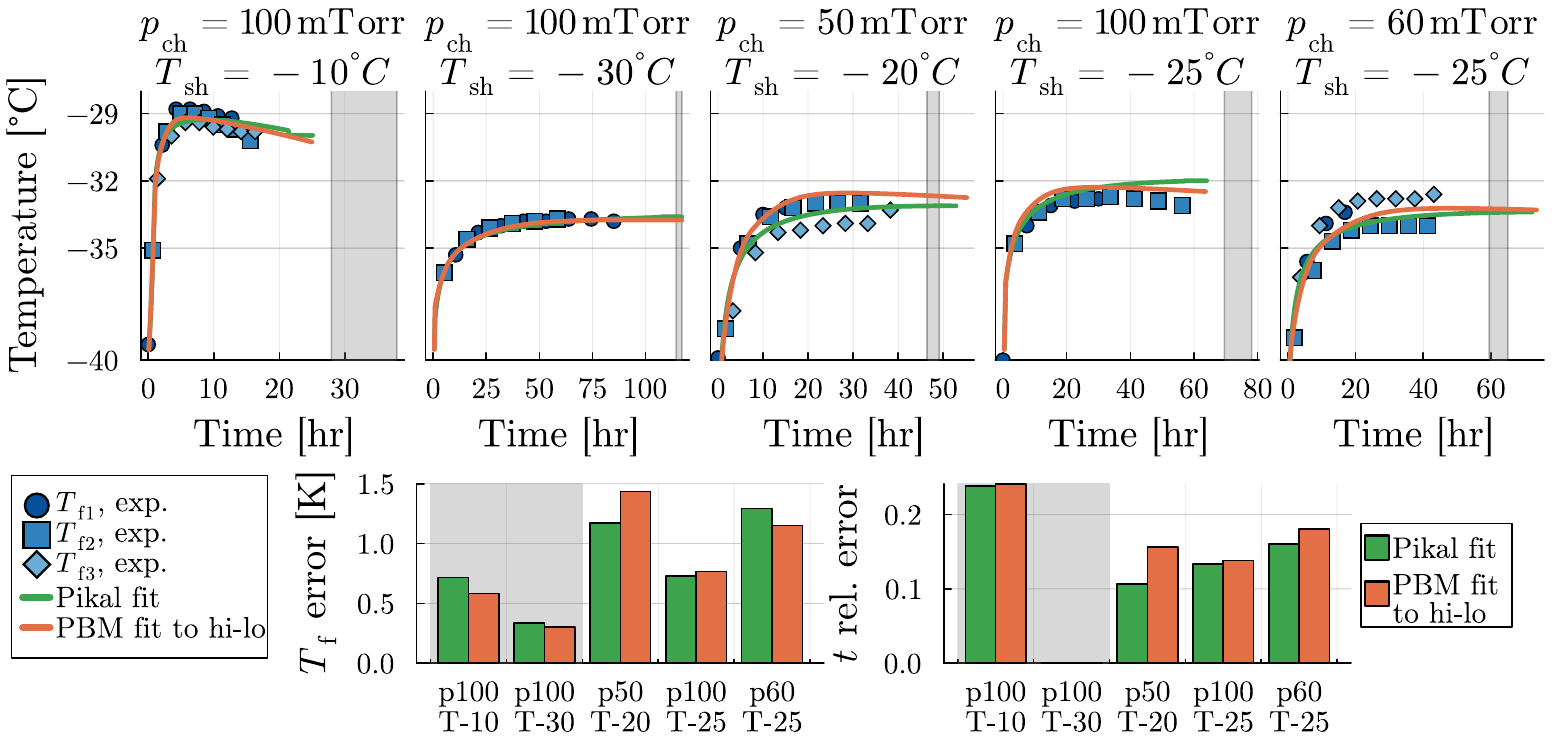}
    \caption{Comparison of in-process thermocouple measurements to model fits for Formulation 1 of \citet{srinivasan_graphical_2023}, where fit is to the two left-most cycles shown and other three cycles serve as validation cases for the fit. Temperature error is the root mean squared error and relative error in time compares total primary drying time, as discussed above in \cref{sec:fits-to-data}}
    \label{fig:simtra-Ts-sucrose}
\end{figure}

\begin{figure}
    \centering
    \includegraphics[width=0.9\linewidth]{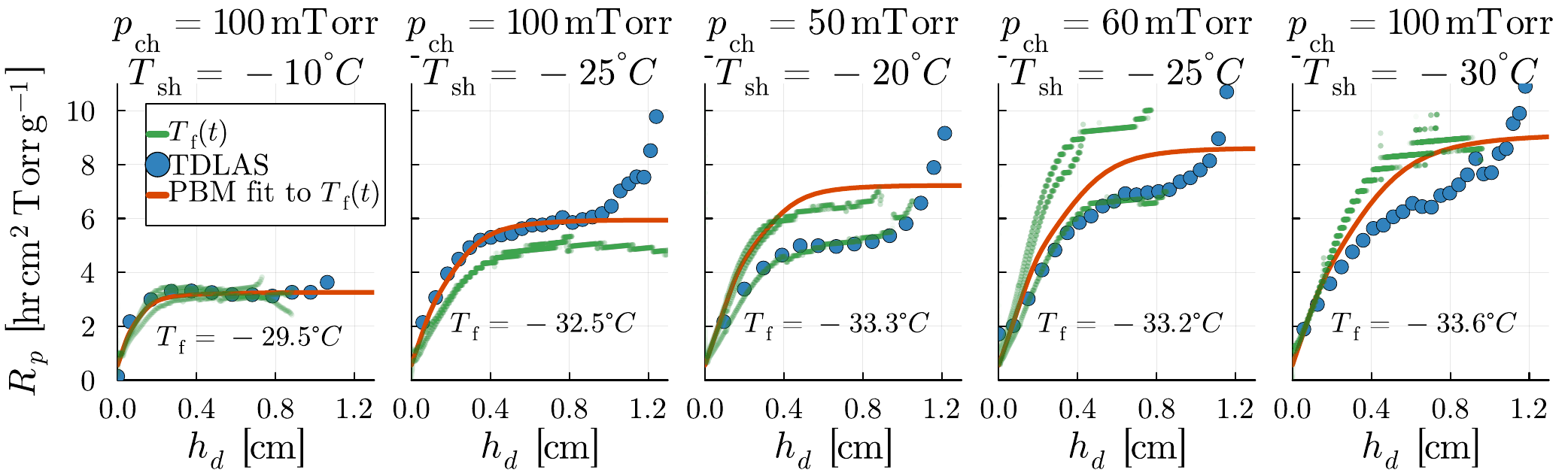}
    \caption{Population balance model fits to two cycles are compared to estimates of $R_p$ from temperature and from TDLAS mass flow. Listed maximum product temperature $T\s{f}$ is taken from the point before temperature rise, as reported in Table 8 of \citet{srinivasan_graphical_2023}.}
    \label{fig:simtra-Rps-sucrose}
\end{figure}

\begin{figure}
    \centering
    \includegraphics[width=0.8\linewidth]{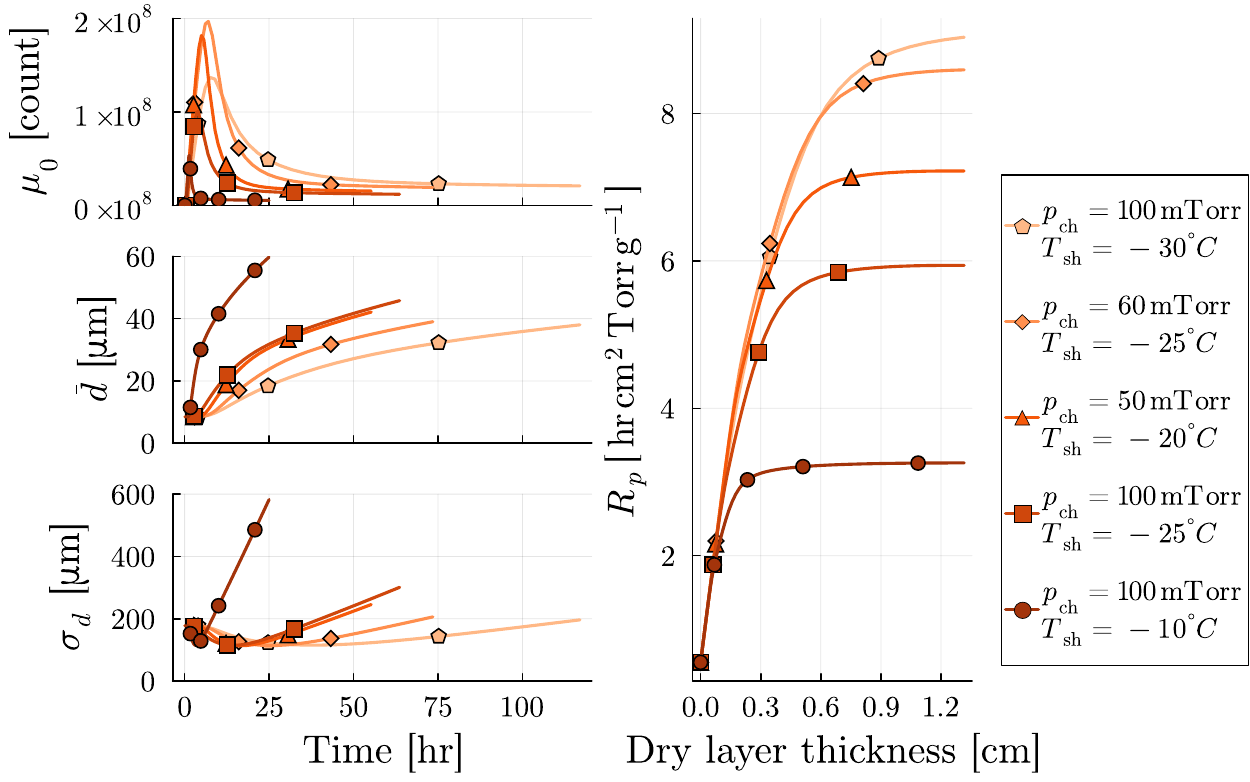}
    \caption{Population balance model results over time for the five examined cases from \citet{srinivasan_graphical_2023}. At left, the number of pores, mean pore size, and pore size standard deviation are shown over time. Cases with smaller $R_p$ (lower in the right panel) complete drying more quickly and this is reflected in the end of the curves in left panels}
    \label{fig:simtra-poresizes-sucrose}
\end{figure}

To investigate variation in $R_p$, \citet{srinivasan_graphical_2023} ran lyophilization cycles at various drying conditions with constant formulation and freezing protocol. The amorphous formulation of 3\% sucrose and 1\% BSA, a simple proxy for many protein formulations, showed significant variation in $R_p$ across drying conditions. Their Table 8 lists process conditions tested for the amorphous formulation, and we reanalyze their process data, as summarized in \cref{tab:srinivasan-cases}. \Cref{fig:simtra-Ts-sucrose} shows the results of fitting the PBM model to their cases 1 and 6 (representing two shelf temperature extremes), and using the same values of fit parameters to evaluate cases 2, 3, and 5 (totaling 8 fit parameters for the 5 cases). 
For comparison, a case-by-case Pikal fit is shown in green (totaling 15 fit parameters across 5 cases); the PBM model fit has very similar accuracy, as shown in the bottom portion of \cref{fig:simtra-Ts-sucrose}. 
In \cref{fig:simtra-Rps-sucrose}, the same simulation results are shown in $h\s{d}$~--~$R_p$ space, where they are compared to experimental estimates of $R_p$ from TDLAS and from thermocouple measurements, as described above in \cref{sec:fits-to-data}.
In \cref{fig:simtra-poresizes-sucrose}, the pore count, mean pore diameter, and standard deviation of pore size are shown for the same simulations.

Although the population balance model usually has slightly more error than the Pikal model when compared to experiment in \cref{fig:simtra-Ts-sucrose}, the difference in accuracy between the two models is small. However, for the Pikal model to match experiment this way, $R_p$ must be fit separately for each case, so it cannot be used at other drying conditions, while the population balance successfully predicts both drying time and product temperatures for conditions where the model was not explicitly fit.
This is strong evidence that the population balance model is successfully describing an underlying physical phenomenon.

This physical description is examined from another direction in \cref{fig:simtra-Rps-sucrose}, where the same five experiments are ordered by ranking maximum $R_p$  in the PBM fit from left to right (so that the two experiments used for fitting are the leftmost and rightmost panes). The blue markers are computed from TDLAS measurements (and so represent a sort of batch-average $R_p$), which is not expected to agree exactly with $R_p$ computed directly from product temperatures (shown by green dots); since the PBM fit is to temperatures, it should fall in a similar range to the $R_p$ computed from temperatures. The rank-ordering by PBM $R_p$ nearly corresponds to the rank-ordering of TDLAS $R_p$ as well as to the maximum product temperatures reported in \citet{srinivasan_graphical_2023}'s Table 8, and falls within the spread of temperature-based $R_p$ for all but the second case from left.
Another piece of the model's output are the moments of the pore size distribution. Formally, the total pore volume $\mu_1$ is already known in the Pikal model; the PBM also computes the number of pores $\mu_0$ and their second moment $\mu_2$, from which we can compute a mean pore volume $ \bar{v} =\mu_1/\mu_0$ and a standard deviation in volumes $\sigma_v =\sqrt{\mu_2/\mu_0 - \mu_1^2/\mu_0^2}$. Both these quantities have dimensions of volume. To aid conceptualization, we relate these to an equivalent spherical diameter by $\bar{d} = \sqrt[3]{\frac{3}{4\pi}\bar{v}}$ and $\sigma_d = \sqrt[3]{2/9\pi}\, \bar{v}^{-2/3}\sigma_v  $. At early times, the number of pores $\mu_0$ increases linearly with drying progress (that is, $\mu_0(t) \approx Nh_d(t)$), but after temperatures rise and microcollapse becomes relevant, $\mu_0$ reaches a steady state at $\frac{k}{2}\mu_0^2 = N\frac{dh\s{d}}{dt}$. The typical pore diameter $\bar{d}$ increases over time, which is expected when microcollapse occurs: at the beginning of drying, all pores match the prior ice crystal sizes, but as pores merge the average size increases.

Although the second moment $\mu_2$ is strictly increasing, the standard deviation may not always increase (i.e. if the mean pore size is growing faster than the second moment). This is observed in \cref{fig:simtra-poresizes-sucrose}, but the standard deviation remains above \qty{90}{\micro\meter} where the mean pore size does not exceed \qty{40}{\micro\meter}. Recalling that, from a mathematical perspective, these moments are exact but the underlying distribution may have any shape, this suggests that the distribution has a long right tail: there are a few pores much larger than the average size, enhancing the vapor flow.

\subsection{Literature Data: Crystalline Formulation with Pore Size Measurements} \label{sec:kosir-dex}

\citet{kosir_impact_2024} used scanning electron microscopy to measure pore size distributions in lyophilized dextran, with varying total concentration and with different shelf freezing conditions, summarized above in \cref{tab:kosir-cases}. Our model here directly includes the first and second moments of the pore size distribution as fit parameters, which can be directly estimated from a measured pore size distribution. If these estimates yield a good fit to experimental data, that suggests that the model is capturing a true underlying physical phenomenon. Additionally, if we can add another tool for measuring these empirical parameters, we can reduce the model's sensitivity to process data and boost confidence in fit parameter values.

\begin{figure}
    \centering
    \includegraphics[width=0.7\linewidth]{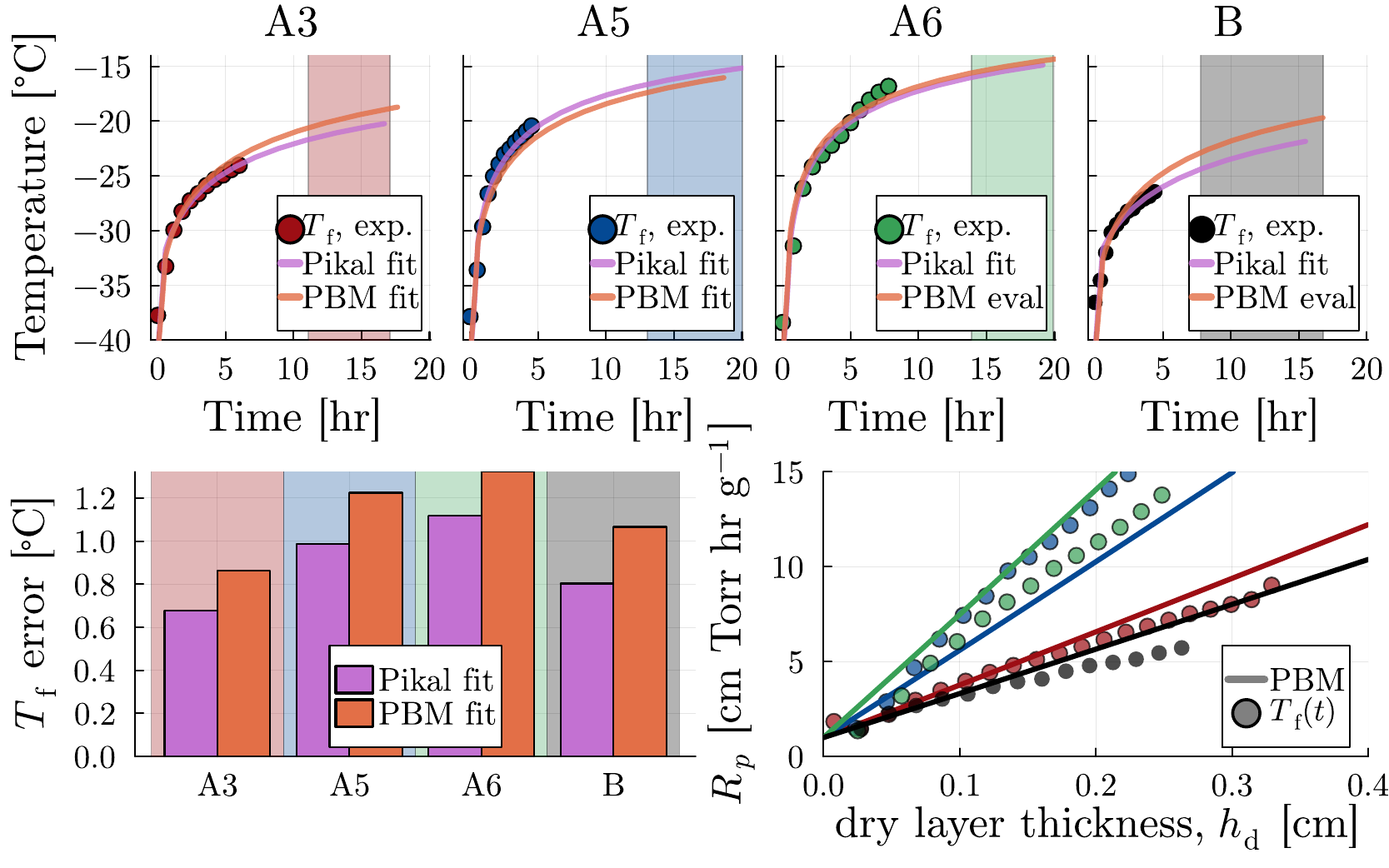}
    \caption{Upper figures: comparison between measured process temperatures, pre-temperature rise, and model fits or predictions for experimental cases in \citet{kosir_impact_2024}, with simultaneous fit of $C$ and $R_s$ to A3 and A5.  Lower left: root-mean-square error in product temperature. Lower right: $R_p$ evaluated with PBM model compared to calculations directly from experimental temperature. Relative error in drying time not shown because all model predictions fall between onset and offset of Pirani convergence (shown as gray shaded region in upper figures), as defined in \cref{sec:exp-postprocess}, so no residual is attributed to time (\cref{eq:time-resid}). }
    \label{fig:kosir-T}
\end{figure}

Dextran is a crystalline excipient and so not prone to microcollapse \cite{tangDesignFreezeDryingProcesses2004}. As discussed in \cref{sec:no-microcollapse}, this allows us to assume $k(T)=\qty{0}{\per\second}$ for these experiments, leaving only $N$, $S$, $C$, and $R_s$ as fit parameters. From temperature data alone, we could not separately fit $N$, $S$, and $C$ (since in the absence of microcollapse, all three parameters are directly multiplied); however, with the pore size distributions $f_D$  shown in Figure~4b of \citet{kosir_impact_2024} (across sphere-equivalent diameter $D$, which can be easily related to volume as used in the present work), we can calculate $N$ and $S$ by
\begin{equation}
\frac{1}{N} = \int_{D=0}^{D\rightarrow \infty} \frac{4\pi}{3}\left(\frac{D}{2}\right)^3f_D dD 
\end{equation}
and 
\begin{equation} 
S = \int_{D=0}^{D\rightarrow \infty} \left[\frac{4\pi}{3}\left(\frac{ D}{2}\right)^3\right]^2 f_D dD .
\end{equation}
Finally, we fit a single value each of $R_s$ and $C$ to temperature measurements from two of their experiments (A3 and A5) and evaluate fit results on another two experiments (A6 and B). These fits are shown in temperature space and in $R_p$ space in \cref{fig:kosir-T}, and the computed values of $N$ and $S$ and fitted values of $C$ and $R_s$ are shown in \cref{tab:kosir-NS}.

\begin{table}
    \centering

    \sisetup{round-mode=figures, round-precision=3, exponent-mode=threshold, exponent-thresholds=-2:3}

\begin{tabular}{@{}rSS@{}}
\toprule
 & {$N$, \unit{\per\milli\liter}} & {$S$, \unit{\milli\liter}}\\
\midrule
{A3} & 4.509859493461617e6 & 4.258150686667547e-7\\
{A5} & 1.4291426651061475e7 & 2.0379879112424856e-7\\
{A6} & 2.0028293799052123e7 & 1.0417073614173452e-7\\
{B} & 2.8452251306546344e6 & 5.852136613902419e-7\\
\bottomrule
\end{tabular}
\\
$ C = \qty{0.18472}{\meter\tothe{1/2}}, R_s = \qty{0.841908}{\hour\centi\meter\squared\torr\per\gram}$

    \caption{Values of $N$ and $S$ extracted from Fig. 3 of \citet{kosir_impact_2024}, with the corresponding fit values of $C$ and $R_s$ which apply to all four cases.}
    \label{tab:kosir-NS}
\end{table}
Since we have applied a single value of $C$ to all four experiments (with $R_s$ taking a small, nonzero value), a good fit between the population balance model and the experimental data implies that the differences in $R_p$ among the four experiments can be explained by the different pore sizes. Indeed, the PBM correctly matches cases A3 and B as having small $R_p$, matching their rank order, and A5 and A6 as having larger $R_p$ (although it mixes up the rank order between these two, which are close together).  The PBM fit (with 2 parameters, across 4 experiments) is also quite close to the case-by-case 3-parameter Pikal fits (with 12 parameters total across 4 experiments). 

\subsection{New Experiments: Crystalline Formulation} \label{sec:ful-dex}

\begin{figure}
    \centering
    \includegraphics[width=0.8\linewidth]{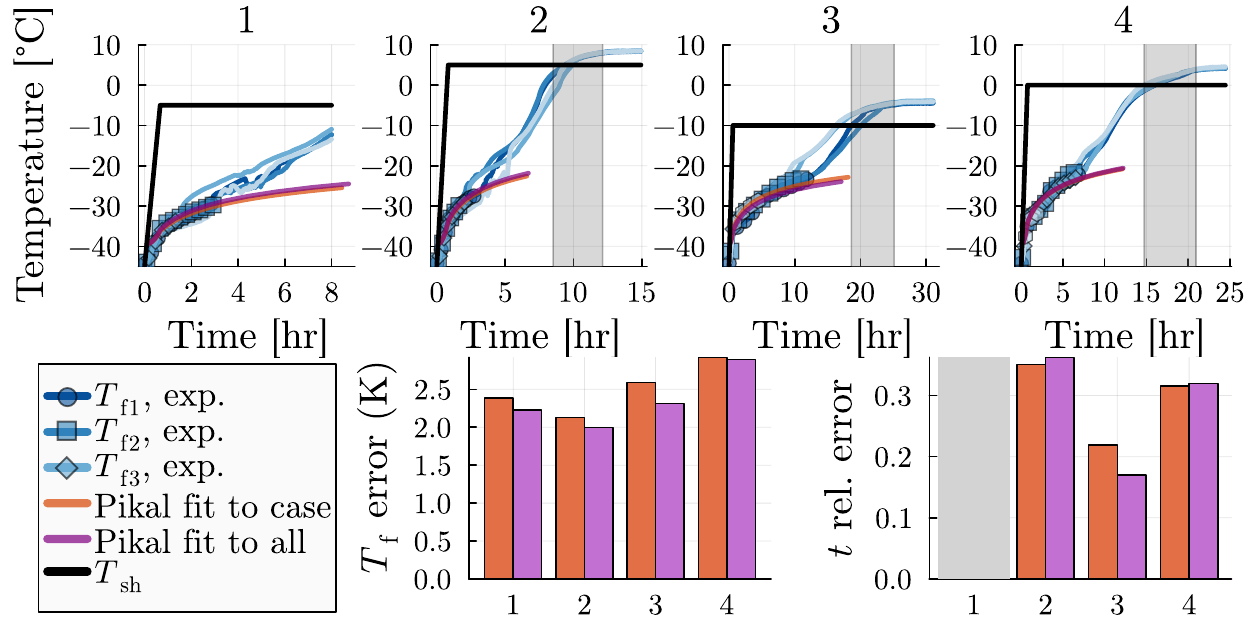}
    \caption{Pikal model fits, either case-by-case or to 4 experiments at once, for a single dextran formulation with single freezing protocol. Data collection for cycle 1 was interrupted, so no end-of-drying information is available.}
    \label{fig:new-dex-T}
\end{figure}

Unlike the cases of \citet{kosir_impact_2024}, if a single formulation is lyophilized without changing the freezing protocol, we can expect approximately similar pore size distributions. For dextran, a crystalline excipient, the lack of microcollapse behavior will lead to similar $R_p$ curves at different drying conditions. Without direct measurements of pore size distribution, the parameters $N$, $S$, and $C$ cannot be distinguished (as discussed in \cref{sec:no-microcollapse}), and so one may as well measure $R_p$ once, fit to the traditional Pikal model, and use it across different drying conditions. This principle is demonstrated in \cref{fig:new-dex-T}: dextran dried at four different drying conditions can be fit nearly as well with a single set of parameters $\{a_0, a_1, a_2\}$ as with separate parameters for each experiment. Since $R_p$ does not have a clear trend as a function of product temperatures, the population balance model does not have any extra insight to offer.

\begin{figure}
    \centering
    \includegraphics[width=0.9\linewidth]{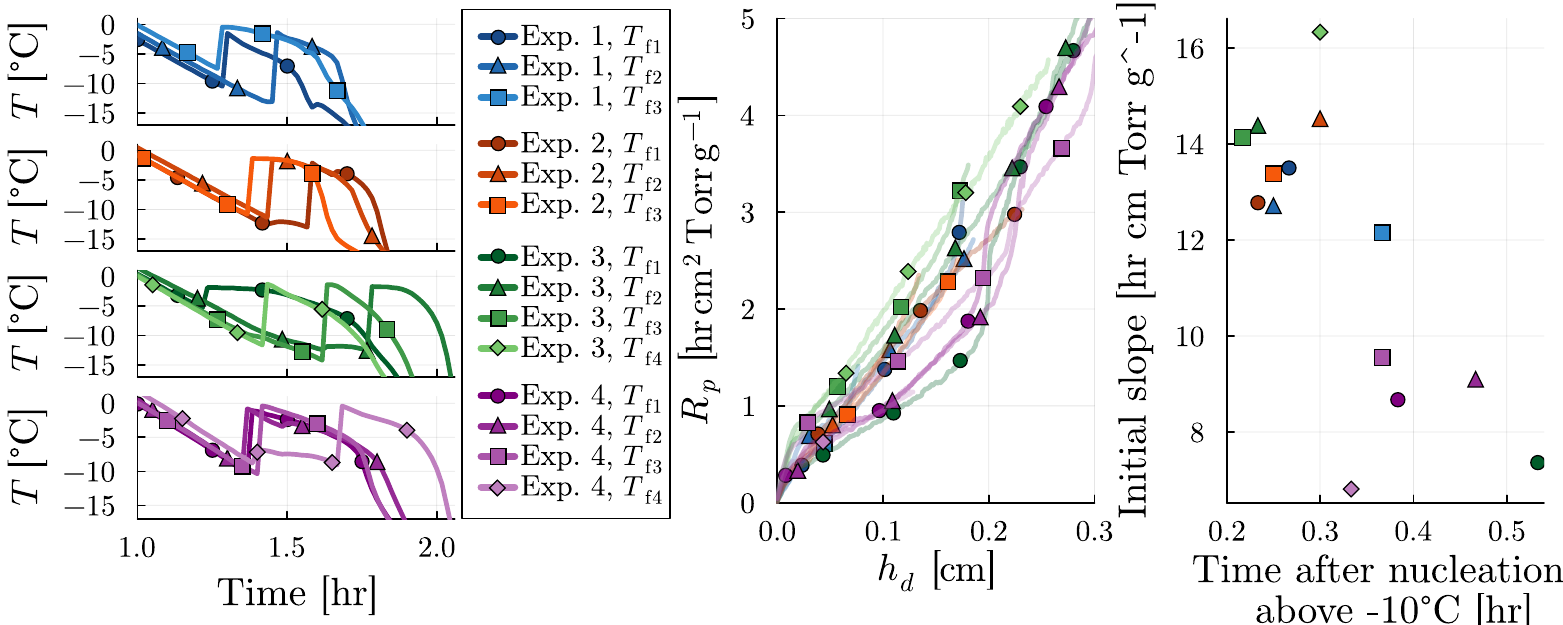}
    \caption{On left: temperatures observed during the portion of freezing when ice nucleation is occurring for 4 cycles of dextran freeze drying. Center: comparison of $R_p$ computed during primary drying from those same thermocouple measurements. Right: a comparison of initial slope in $R_p$ against the time from ice nucleation to temperature descending below \qty{10}{\celsius} (called the residence time by  \citet{kharatyan_rethinking_2026}).}
    \label{fig:new-dex-Rp}
\end{figure}

To suggest a possible direction for future work tying ice crystal size distributions to freezing behavior (which then would provide estimates of $N$ and $S$), 
we analyze the freezing and nucleation behavior following \citet{kharatyan_rethinking_2026}. They propose that the several-minute-long period after ice crystal nucleation and before complete solidification, while the product is near freezing temperature, functions like an annealing step and causes Ostwald ripening of ice crystals immediately after nucleation. This is assessed in \cref{fig:new-dex-Rp}, which shows 1) the temperature in the freezing step, 2) the computed $R_p$ (i.e., plugging $T\s{f}(t)$ into the Pikal model and evaluating $R_p(t)$ and $h\s{d}(t)$ over time) for all 16 thermocouple traces, and 3) the initial slope of $R_p$ vs. the time spent above \qty{-10}{\celsius}. Recall that when the parameter $a_2$ in \cref{eq:pikal-rp} is zero, the slope of $R_p$ is simply $a_1$, which may be related to permeability of the porous structure \cite{kosir_elucidating_2025, kuu_product_2011}.  Across all the experiments, $R_p$ seems to fall on a single line above $h\s{d}=\qty{0.2}{\centi\meter}$, but some have a different slope initially, and this initial slope of $R_p$ is the vertical axis of the right panel.  

There seems to be a decreasing correlation in the right of \cref{fig:new-dex-Rp}, which supports the analysis of \citet{kharatyan_rethinking_2026} and suggests that more time for ice crystal growth is translating to larger ice crystals and larger pores, perhaps in a way which can be quantitatively predicted with further development. However, two features complicate the story: first, there are outliers both above and below the apparent line; second, if the overall average slope were used rather than initial slope, the points would scatter in a roughly vertical shape.

\subsection{New Experiments: Amorphous Formulation} \label{sec:ful-suc}
To assess whether the microcollapse rate constant $k(T)$ and tortuosity parameter $C$ may be thought of as constant for a given material, irrespective of freezing conditions and formulation concentration, we attempt to fit 6 sucrose lyophilization experiments simultaneously, with varying sucrose concentration and one experiment using controlled ice nucleation (CIN). For a given formulation (i.e., concentration) and freezing condition, we expect to reproduce approximately the same ice crystal size distribution, and so the parameters $N$ and $S$ (zeroth and second moments of ice crystal size distribution) are held constant across primary drying temperature but not across freezing and formulations. $C$ and $k(T)$ are held the same across all conditions. Thus, across 6 experiments with sucrose, we have 12 total fit parameters:
\begin{enumerate}
    \item 3 values each of $N$ and $S$ (5\% with CIN, 5\% without CIN, 10\% without CIN): 6 parameters
    \item 1 value each of $C$ and $R_s$: 2 parameters
    \item 1 value each for $k_0$, $k_f$, $K_1$, $K_2$ to compose $k(T)$: 4 parameters
\end{enumerate}

For this set of experiments, the time-vs-temperature weighting in the fitting loss function was reduced to $w=\qty{0.01}{\kelvin\squared\per\hour\squared}$ (see \cref{sec:fits-to-data} for definition), which places higher emphasis on matching temperature variation and less on matching drying time.

\begin{figure}
    \centering
    \includegraphics[width=0.9\linewidth]{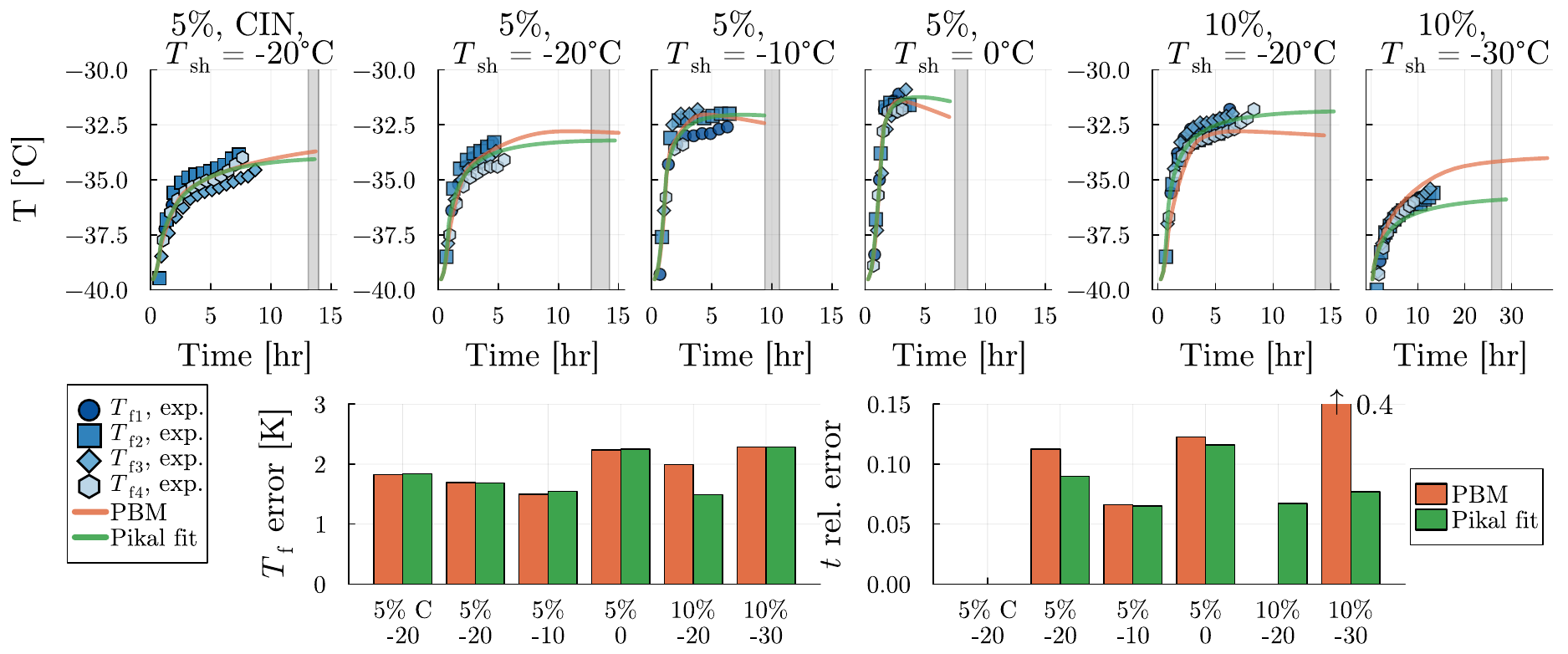}
    \caption{PBM fit to 6 sucrose lyophilization experiments, simultaneously. $C$, $R_s$, and $k(T)$ are shared across all cases; $N$ and $S$ are each given 3 distinct values, one for the 5\% sucrose with controlled ice nucleation (CIN), one for 5\% sucrose, and one for 10\% sucrose. Relative error in time for one PBM model prediction far exceeds the rest, which is marked on the bar chart with an arrow.}
    \label{fig:ful-suc-T}
\end{figure}

\begin{figure}
    \centering
    \includegraphics[width=0.9\linewidth]{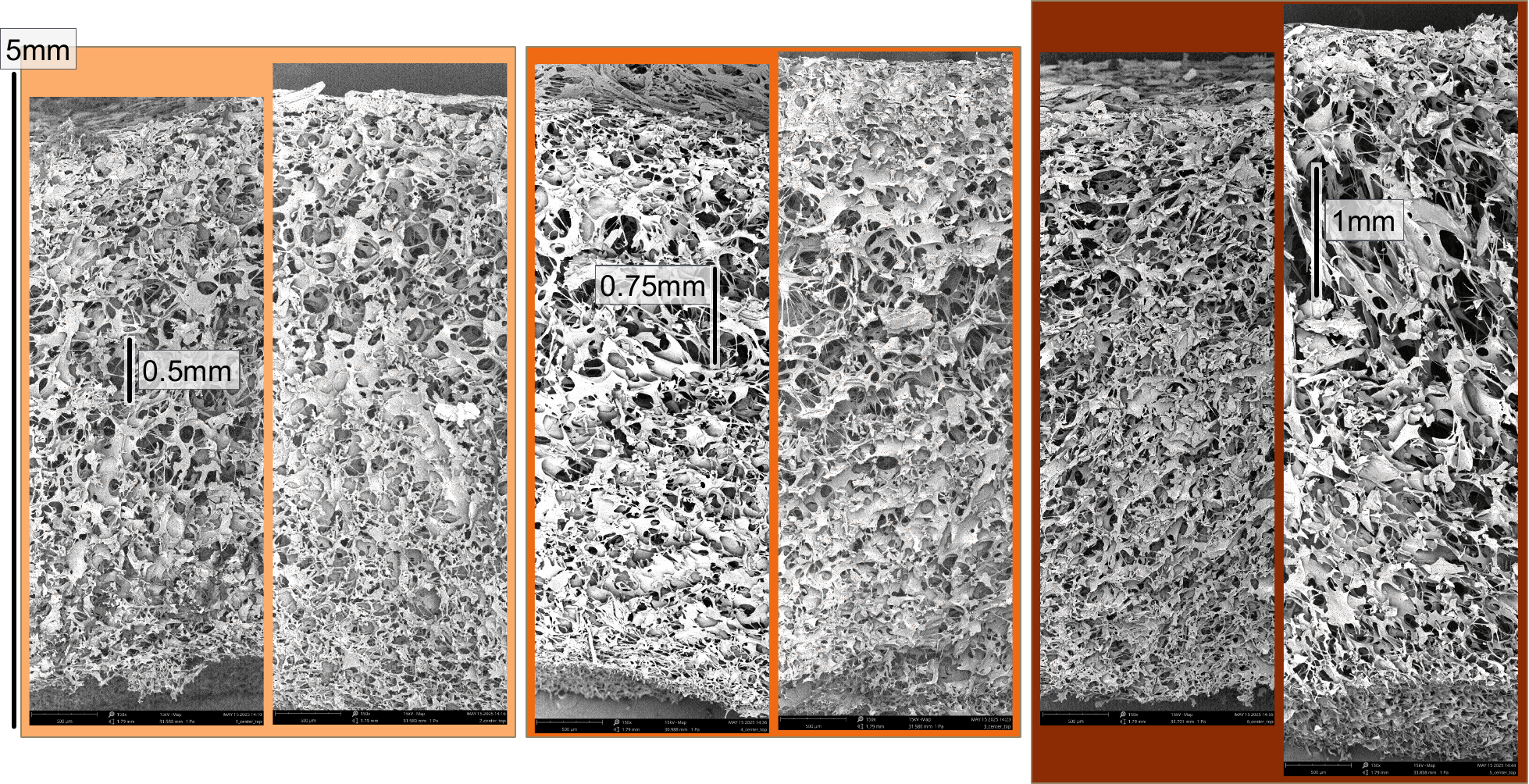}
    \caption{Composite SEM images of 5\% sucrose at common scale (with scale bars highlighting pore sizes), with identical freezing protocol and lyophilized at shelf temperatures of -20, -10, and \qty{0}{\celsius} respectively. Each image shows the full height of a lyophilized cake, sliced and imaged near the center radially, and is composed of 5-10 separate SEM images each having a 1.8mm field of view. }
    \label{fig:ful-suc-SEM}
\end{figure}

\begin{figure}
    \centering
    \includegraphics[width=0.9\linewidth]{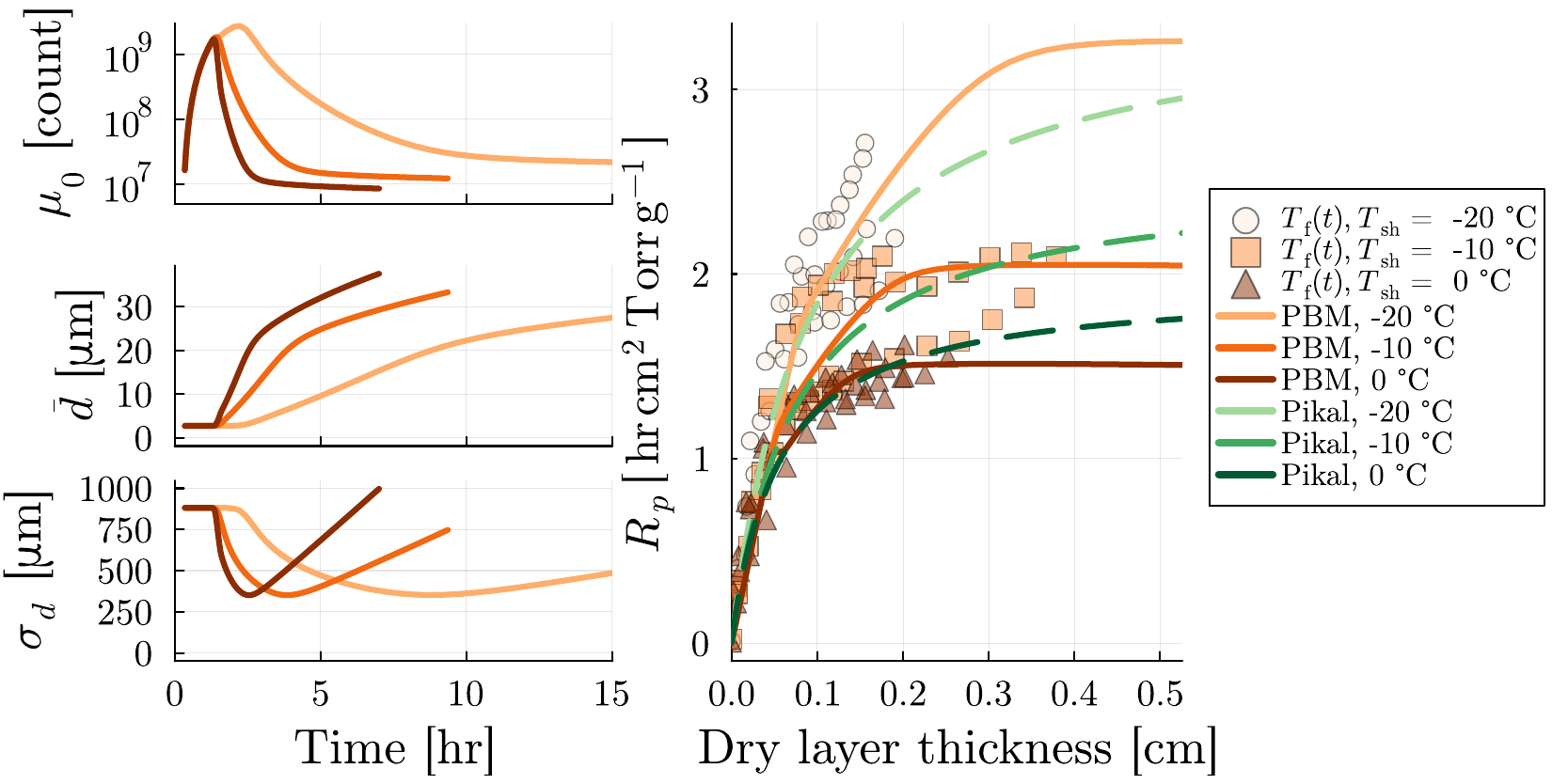}
    \caption{Left: evolving moments of the pore size distribution for model fit to 5\% sucrose experiments with shelf temperature of -20, -10, and \qty{0}{\celsius}, corresponding to the three experiments for which SEM images are shown in \cref{fig:ful-suc-SEM}. Right: $R_p$ from population balance model and computed directly from temperatures for the same three cases.}
    \label{fig:ful-suc-mom}
\end{figure}

The final fit results are shown in \cref{fig:ful-suc-T}, compared to a case-by-case Pikal fit (with $6\times 3 = 18$ parameters). For the 5\% sucrose formulation without CIN the resulting pore moments are shown in \cref{fig:ful-suc-mom} which can be compared to SEM images in \cref{fig:ful-suc-SEM}.

Broadly, the PBM fit results in \cref{fig:ful-suc-T} are of similar accuracy to the Pikal fit, although predicted product temperatures at the end of drying (when no experimental measurement is available) are different between the two models. The major exception are the cases with 10\% sucrose, where the PBM fit predicts lower or higher product temperatures respectively and significantly overestimates drying time for the $T\s{sh}=\qty{30}{\celsius}$ (predicting \qty{37}{\hour} for an experimental drying time of \qty{26}{\hour}). Recalling that all experiments were present and weighted equally in the parameter fit, this suggests that $k(T)$ may not be solely a function of the freeze-concentrate's composition (which is the same across all 6 experiments), or that some other confounding factor affected the one experiment. 

\Cref{fig:ful-suc-T} also serves as a clear example of the importance of including the primary drying time in model fitting procedure. Consider in particular the rightmost case, with 10\% sucrose at $T\s{sh}=\qty{-30}{\celsius}$: at such low product temperatures, we might expect to see no microcollapse and therefore have $R_p$ and $T\s{f}$ both increasing in the time near end of drying when product temperatures are unavailable. However, this would reduce the net heat transfer (i.e. $\int_0^t K_vA_v(T\s{sh}-T\s{f}$), and lead to model predictions of drying time much longer than the measured drying time of \qtyrange{25}{27}{\hour}, so even the case-specific Pikal model predicts a nearly-flat product temperature.

Future exploration could consider fitting the PBM to case-by-case Pikal model fits. This two-step fit would provide a natural way of imposing existing intuition about the behavior of the process, particularly at the end of drying when temperature measurements become unreliable.

The porous structures in \cref{fig:ful-suc-SEM} cannot be easily segmented into individual pores. Even the criteria established by \citet{kosir_impact_2024} require significant subjective application here: for example, very few pores in the exposed layer here have a fully visible wall structure. Qualitatively, at least, the three structures do not seem especially different, even though fitted $R_p$ varies by a factor of 2 at the end of drying. This is corroborated by the mean pore diameters, which change only from \qtyrange{65}{72}{\micro\meter}. However, the largest connected spaces highlighted by the additional scale bars on the 1st, 3rd, and 6th images do increase in size from about \qty{0.5}{\milli\meter} to about \qty{1}{\milli\meter}, which agrees with the trend in standard deviation shown in the lower left of \cref{fig:ful-suc-mom} which is about \qty{85}{\micro\meter} for the \qty{-20}{\celsius} case and \qty{115}{\micro\meter} in the \qty{0}{\celsius} case (at end of drying). The model is somewhat insensitive to the exact values of fitted pore sizes, so long as the second moment is adjusted to compensate; depending on initial guess, similar quality of fit could be obtained with pore sizes as small as \qty{10}{\micro\meter} or standard deviations as large as \qty{500}{\micro\meter}, although these alternative fits are not shown here. In general, however, the population balance model fit is able to match the trend in $R_p$ that the Pikal model fits produce.  

\section{Discussion and Future Directions} \label{sec:discussion}

\subsection{Utility of the Population Balance Model} \label{sec:utility}

To construct a design space for the lyophilization process, like those demonstrated by \citet{srinivasan_graphical_2023} or \citet{vanbillemont_optimization_2025}, the mass transfer resistance $R_p$ is crucial. As highlighted by \citet{srinivasan_graphical_2023} (with results likewise shown above in \cref{fig:simtra-Rps-sucrose}), for amorphous materials the mass transfer resistance varies inversely with the observed product temperatures during drying. As a consequence, low-temperature measurement of $R_p$ will place a stricter constraint on the design space than a high-temperature measurement of $R_p$. Since higher product temperatures and lower $R_p$ both increase the drying rate (all else being equal), constructing the design space with a measurement of $R_p$ which more closely matches behavior near the upper limit of temperature will make apparent that the product can safely be lyophilized in a region of the design space where sublimation rate is higher.

The population balance model we present here provides a way to more rigorously create such a design space: rather than iteratively measuring $R_p$ at higher and higher temperatures with lower and lower $R_p$, one experimental measurement each at a low and a high temperature can allow interpolation and perhaps even extrapolation.
The algorithmic construction of a design space utilizing these two measurements is an important future direction for this work.

\subsection{Temperature Dependence of Microcollapse Rate Constant} \label{sec:results-temp-dep}

Since the composition of the freeze concentrate is a thermodynamic quantity set by equilibrium between the freeze concentrate and ice \cite{zah_predicting_2023}, a 5\%, a 10\%, or a 20\% solution will all produce the same freeze concentrate composition, so long as the molar ratios of non-water species (e.g. 1 part protein to 4 parts sucrose) remain constant. The rigidity or viscosity of the resulting amorphous solids is thus not dependent on the overall solids content of the formulation.
If in like manner the microcollapse rate constant $k(T)$ is the same for both a 5\% sucrose formulation and a 10\% sucrose formulation , this would be one piece of evidence suggesting that $k(T)$ is directly tied to the material parameters (like rigidity or viscosity) of the freeze-concentrated sucrose material, and $k(T)$ might then be separately measurable by mechanical characterization of the freeze concentrate. 

\begin{figure}
    \centering
    \includegraphics[width=0.8\linewidth]{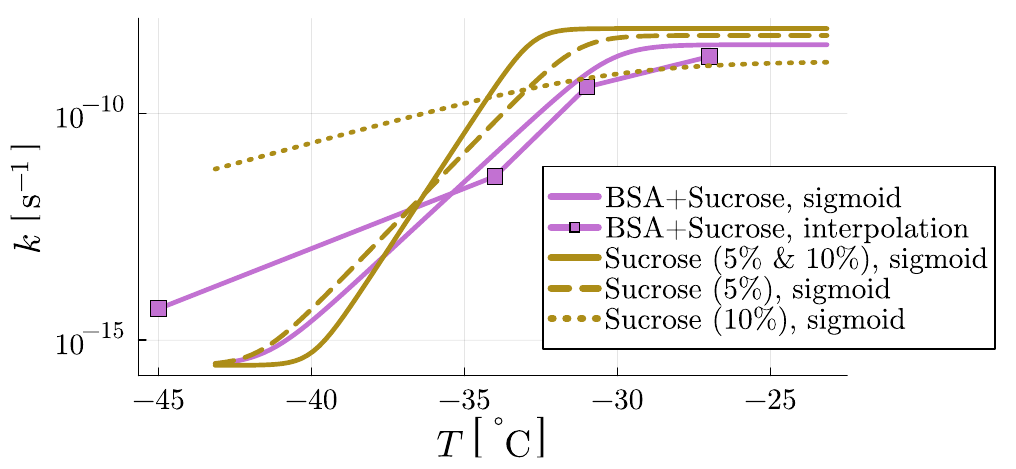}
    \caption{Best-fit rate constants, with sigmoid functional form (or interpolative fit with markers) for the population balance fit to experiments shown in \cref{fig:simtra-Ts-sucrose} and \cref{fig:ful-suc-T}.}
    \label{fig:rate-constants}
\end{figure}

In \cref{fig:rate-constants}, we present the fitted forms of $k(T)$ which underlie \cref{fig:ful-suc-T} and \cref{fig:simtra-Ts-sucrose}, alongside some hypothetical alternative fits. 
The sigmoid fits all collapse onto a single value of $k$ at low $T$, which is an artifact of the fitting process: for the examined cases, any values of $k$ less than about \qty{1e-13}{\per\second} lead to essentially no microcollapse, so the fitting procedure cannot tell apart values below that threshold and does not noticeably adjust the lower-bound $k_0$ away from the guess value of about \qty{3.5e-16}{\per\second}.  

We attempted to assess above in \cref{sec:ful-suc} whether $k(T)$ is independent of solids content for a pure sucrose formulation. However, this is somewhat difficult to rigorously determine with the present experiments. 
A fit only to experimental temperatures, as shown in \cref{fig:ful-suc-T}, admits a potentially broad range of microcollapse rate constants---the combination fit to both 5\% and 10\% shown here in \cref{fig:rate-constants} results in very similar experimental error to the separate fits of 5\% and 10\%, despite apparent differences in the rate constant. For this reason, we caution against overinterpretation of the exact values of the rate constant and its functional forms shown here.

Across the temperature range from \qtyrange{-45}{-25}{\celsius}, an Arrhenius fit with maximum-possible slope (which occurs at activation energy equal to $R_uT$ for a temperature $T$ of interest) can increase at most by about a factor of 2, where even the most-conservative fit shown here (the 10\% sucrose sigmoid fit) increases by 3 orders of magnitude, so this microcollapse rate constant cannot be reasonably described by a typical Arrhenius fit.

In multi-experiment fits as used for \cref{sec:simtra-suc} and \cref{sec:ful-suc}, the sigmoid function \cref{eq:sigmoid} described above in \cref{sec:model-temp-dep} tends to be much more better-behaved in the fitting process. Although the MYEGA form \cref{eq:MYEGA} might provide more physical interpretability, it is very sensitive to the innermost fit parameter $K_2$, and this causes practical difficulties in fitting model parameters to experimental data. In model development, we explored an alternative approach of fitting $k(T)$: for a set of four temperatures $T_1, T_2, T_3, T_4$ strategically chosen in the range of \qtyrange{-40}{-28}{\celsius}, fitting determines four values $k_1=k(T_1)$, etc., and $k(T)$ is evaluated by interpolating with $\ln (k/k_0) = \frac{\ln(k_{i+1}/k_0) - \ln(k_i/k_0)}{T_{i+1}-T_i}(T-T_i)$ for $T_i < T < T_{i+1}$ and some reference $k_0$. This provides a way to assess the approximate temperature dependence of $k(T)$ without assuming a functional form, and the values $k_1, k_2, k_3, k_4$ can be used to estimate appropriate values of the fit parameters $k\s{ref}$, $K_1$, and $K_2$ in the MYEGA equation. However, even considering this two-step fit, the sigmoid form proved more numerically tractable and thus yielded better fits, which is why it is used in all the results we present above.

\subsection{Flow Rarefaction} \label{sec:Kn}
\Cref{eq:Rp-knudsen}, which relates moments of the pore size distribution to $R_p$, is derived with the assumption of Knudsen diffusion through a bundle of capillaries. The assumption of rarefied Knudsen flow behavior is reasonable when the Knudsen number $\mathrm{Kn}$, equal to a ratio of the gas mean free path and characteristic length, is large; that is, for $\mathrm{Kn}>0.1$, flow is in a transitional regime and for $\mathrm{Kn}>1$ flow is in a fully rarefied regime, whereas the no-slip viscous flow of the Hagen-Poiseuille law occurs for $\mathrm{Kn}<0.001$.  

\begin{figure}
    \centering
    \includegraphics[width=0.5\linewidth]{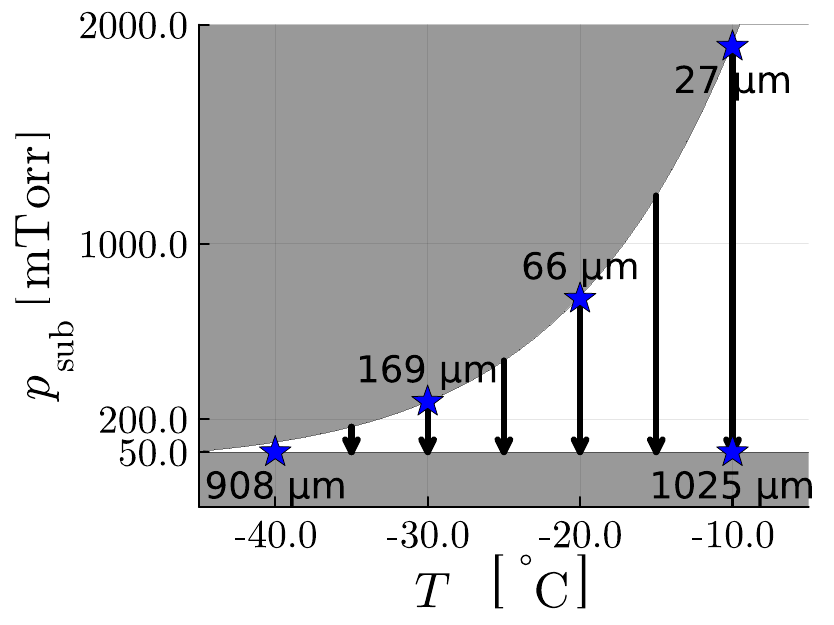}
    \caption{Typical temperatures and pressures of vapor flow in the dried cake during primary drying, annotated with corresponding mean free path of water vapor at those conditions. Curved upper limit corresponds to equilibrium vapor pressure over ice as a function of temperature; flat lower limit of \qty{50}{\milli\torr} corresponds to typical chamber pressure. Downward arrows visually suggest that pressure drops along the flow path from sublimation front to chamber}
    \label{fig:typical-Kn}
\end{figure}

We can estimate the mean free path of the water vapor as 
\begin{equation}
    \Lambda = \frac{k_BT}{\sqrt{2}\pi d\s{ref}^2 p}
\end{equation}
where $k_B$ is the Boltzmann constant and $d\s{ref}$ is an effective hard-sphere molecular diameter, taken to be \qty{3.46}{\angstrom} for water vapor as in \citet{kuu_product_2011}. The mean free path is especially sensitive to pressure $p$ ; at typical freeze drying conditions, $T$ varies only by 10\% (that is, \qty{25}{\kelvin} variation around \qty{250}{\kelvin}), but the vapor pressure over ice in equilibrium with these temperatures changes by orders of magnitude. For a chamber pressure of \qty{50}{\milli\torr} (as is common at the lab scale), the pressure trajectory for flow from sublimation front to chamber is schematically indicated in \cref{fig:typical-Kn}. Also shown are the mean free paths calculated at several points in the space of typical conditions, which is bounded above by the sublimation pressure and below by chamber pressure.

To estimate the characteristic length scale, we use pore diameters. The largest pores observed by \citet{kosir_impact_2024} are around \qty{150}{\micro\meter}, with average diameter closer to \qty{50}{\micro\meter} (see their Fig. 4b). For the amorphous formulation of \citet{srinivasan_graphical_2023}, the above model fit gives average pore diameters no larger than \qty{30}{\micro\meter}, and standard deviations no larger than than \qty{130}{\micro\meter} (see \cref{fig:simtra-poresizes-sucrose}); very large pores thus may be \qty{300}{\micro\meter} in diameter. Literature measurements of pore sizes fall in this same range, such as \citet{parker_determination_2010} (their Fig. 3, \qtyrange{4}{150}{\micro\meter}), \citet{izutsu_studying_2014} (their Fig. 4, \qtyrange{50}{100}{\micro\meter}) and  \citet{pisano_characterization_2017} (their Fig. 3 \& Fig. 5, \qtyrange{10}{150}{\micro\meter}).

So, for especially large pores close to the sublimation front at high temperatures (i.e. $T\s{sub}=\qty{-10}{\celsius}$), the Knudsen number $\mathrm{Kn}=\Lambda/L$ may reach as small as $\qty{27}{\micro\meter} / \qty{300}{\micro\meter} \approx 0.1$. This lower bound coincides with the conventional lower bound of the transitional flow regime, so we should not expect viscous flow with or without slip conditions. For small pores at low temperatures we may reach $\qty{1000}{\micro\meter} / \qty{10}{\micro\meter} = 100$, an upper bound which is above the upper limit of the transition regime at $\mathrm{Kn}=10$. This upper bound estimate assumes small pores, which will lead to very slow drying, so these conditions will often be avoided in practice. Therefore the bulk of flow behavior will be in the transition regime, in which case the Knudsen diffusion expression is unlikely to be fully correct from a fluid-mechanics perspective; this should be investigated in future work.

\section{Conclusions}

We present here a novel vial-scale model for mass transfer during the primary drying step of lyophilization.
This model employs the heat transfer portion of the literature-standard model \cite{tchessalov_practical_2023} as-is, and rewrites completely the mass transfer component of that model: we replace the weakly nonlinear empirical expression for mass transfer resistance (\cref{eq:pikal-rp}) with an expression in terms of moments of the pore size distribution (\cref{eq:Rp-knudsen}), and add equations that describe time evolution of these moments (\cref{eq:mu0-post}, \cref{eq:mu2-post}). 

This population balance on pore space yields a mass transfer model which achieves two key goals:
\begin{enumerate}
    \item  In amorphous material (i.e. susceptible to microcollapse during lyophilization) and with uniform formulation and freezing, the population balance model can predict mass transfer resistance across different drying conditions, so long as it is fit to at least one conservative and one aggressive case.
    \item With separate measurements of (the moments of) pore size distributions in a non-microcollapsing material, the population balance model is capable of matching several experiments with a single value of the tortuosity-like parameter $C$. 
\end{enumerate}

Future work can establish a rigorous functional form for the microcollapse rate constant $k$ as a function of temperature, as well as explore a more rigorous expression relating moments of the pore size distribution to $R_p$.

With this model, we anticipate that researchers and engineers can reduce the number of experiments necessary to design optimal lyophilization cycles for amorphous excipients. This model may also facilitate cycles which, with appropriate validation of product critical quality attributes, deliberately trigger microcollapse in order to cut cycle duration and produce more doses of lifesaving medicines.
\vspace{6pt}

\section{Declarations}
\subsection{Author contributions}
Conceptualization and methodology, I.S.W. and D.F.; software, validation, formal analysis, investigation, data curation, writing---original draft preparation, project administration and funding acquisition, I.S.W.; resources, D.F.; writing---review and editing, all authors; visualization, I.S.W.; supervision, D.F., A.A.A., and V.N.  All authors have read and agreed to the published version of the manuscript.

\subsection{Funding}
I.S.W.'s visit to the Politecnico di Torino was partially funded by a Fulbright grant. There are no other funding sources to report.

\subsection{Data availability}
The scripts and analyzed data are available in a Zenodo repository at \cite{isaacwheelerIckaserFulbrightPorousjlV0102026} which archives releases from a GitHub repository at \hyperlink{https://github.com/Ickaser/FulbrightPorous.jl}{https://github.com/Ickaser/FulbrightPorous.jl}. By including a \verb|Manifest.toml| file, the full tree of dependencies for the Julia code is recorded.

\subsection{Acknowledgements}

Fiora Artusio and Roberto Pisano provided guidance in the preparation of samples for SEM imaging. Francesco Fiore operated the SEM alongside I.S.W. . 

During the execution of this study, I.S.W. used line-level autocomplete from GitHub Copilot to assist in the iterative process of data analysis. All code and text was otherwise written without the use of generative AI tools.

\subsection{Conflicts of Interest}

A.A.A. holds equity ownership in LyoWave, Inc which commercializes equipment for a microwave-assisted lyophilization process. I.S.W., D.F, and V.N. declare no competing interests.

\appendix
\section{Mass Flux in Capillaries} \label{sec:app-massflux}

The mass flux for Knudsen diffusion in a capillary of length $l$ is
\begin{equation}
    J_K = D_K M_W\nabla C 
    = \left( \frac{2r}{3} \sqrt{\frac{8R_uT}{\pi M_W}}\right) M_W \left( \frac{\Delta p}{R_u T l} \right) 
    = \frac{4\sqrt{2}}{3\sqrt{\pi}} \sqrt{\frac{M_W}{R_uT}} \frac{r\Delta p}{l}
    \label{eq:JK}
\end{equation}
where the capillary diameter $2r$ is being used to estimate the mean free path.
For viscous flow through a capillary (or tube) of radius $r$, the Hagen-Poiseuille law relates pressure drop $\Delta p$ to volumetric flow $\dot{V}$ by
\begin{equation}
    \Delta p = \frac{8 \eta l \dot{V}}{\pi r^4}
    \label{eq:hagen-poiseuille}
\end{equation}
with $\eta$ the dynamic viscosity.
To rewrite this as a flux like $J_K$ above, we let $J_V= \dot{V}\rho/\pi r^2$  and evaluate the density $\rho$ with the ideal gas law: 
\begin{equation}
    J_V = \frac{\Delta p r^2 \rho}{8 \eta l} = \frac{\Delta p r^2 M_Wp}{8\eta lR_uT} .
    \label{eq:JV}
\end{equation}
To put these fluxes in the form of resistances, this flux is evaluated for a given radius, then integrated across a distribution of radii corresponding to an imagined bundle of capillaries to evaluate $\dot{m}$, as in \cref{eq:mass-from-flux}. Finally, rearrange this expression to be in the form of $R_p= A_p \Delta p/\dot{m}$ to find \cref{eq:Rp-knudsen} and \cref{eq:Rp-viscous}.

\bibliographystyle{unsrtnat}
\bibliography{references}

\end{document}